\documentclass[final,3p,times,number]{elsarticle}

\usepackage[hyphens]{url}   
\usepackage{float}
\usepackage[section]{placeins}
\usepackage{amsmath}
\usepackage{amssymb}
\usepackage{booktabs}
\usepackage{caption}       
\usepackage{subcaption}    
\usepackage{float}
\usepackage[hidelinks]{hyperref}              
\usepackage[capitalize,nameinlink]{cleveref}  

\def\tsc#1{\csdef{#1}{\textsc{\lowercase{#1}}\xspace}}
\tsc{WGM}
\tsc{QE}
\tsc{EP}
\tsc{PMS}
\tsc{BEC}
\tsc{DE}

\begin{document}

\begin{frontmatter}

\title{Understanding the role and value of battery storage in renewables-rich distribution grids under uncertainty and different market paradigms}

\author[1,2]{Peer Valentin Brigger}

\author[1]{Vineet Jagadeesan Nair}

\author[1]{Anuradha~Annaswamy\corref{cor1}}
\ead{aanna@mit.edu}

\cortext[cor1]{Corresponding author}

\address[1]{Department of Mechanical Engineering,
  Massachusetts Institute of Technology,
  Cambridge, MA 02139, USA}

\address[2]{Department of Information Technology and Electrical Engineering, ETH Zurich, 8092 Zurich, Switzerland}

\begin{abstract}
Economy-wide electrification is driving steep load growth while the rapidly increasing penetrations of renewables and batteries are creating pressure at the distribution grid edge, often necessitating costly grid upgrades. Local Electricity Markets (LEMs) have been shown to coordinate distributed energy resources (DERs) more efficiently than conventional retail designs. This paper extends the LEM framework by modeling photovoltaic (PV) inflow as an Ornstein–Uhlenbeck (OU) process to capture realistic forecast uncertainty and temporal correlation. Using a rolling-horizon alternating optimal power flow on the IEEE 123-node test feeder, operator and customer costs are compared as well as system costs, and storage valuation under three market designs: flat tariff (FT), real-time pricing (RTP), and LEM. Results show that the LEM achieves the lowest customer and system costs, even under uncertainty, while RTP performs the worst. The LEM significantly outperforms the other markets across most penetration levels of storage and solar. However, it may worsen outcomes for customers in networks with very low PV penetration -- indicating that LEM adoption must be accompanied by PV expansion and that PV adoption needs to be prioritized before expanding storage capacity. We explore two different approaches for estimating battery storage value. We find that the locational and temporal value of storage depends strongly on several factors such as the overall storage penetration level, imbalance between storage and PV penetrations, storage locations, and whether batteries are colocated with renewables and/or load at the same node. Poorly placed units can worsen congestion, whereas well-sited ones deliver significant benefits to both the owner and the system. We find that both the marginal value of new storage and the stored energy value of existing capacity decline with increasing PV penetration. However, they respond oppositely to battery penetration: marginal value falls due to diminishing returns as the fleet grows, while stored energy value rises as more batteries compete for the same local PV generation. These two metrics thus capture fundamentally different dimensions of storage value under a LEM -- one measures the benefit of expanding the fleet, while the other measures the scarcity premium on energy already inside it.
\end{abstract}

\begin{highlights}
\item A local electricity market eliminates lost load and lowers system cost under PV forecast uncertainty, outperforming flat-tariff and real-time pricing designs.
\item The LEM delivers better average performance but amplifies sensitivity to PV forecast uncertainty relative to benchmark designs.
\item LEM performance is threshold-driven: it strongly outperforms benchmarks above a $\sim$15--20\% PV penetration but can worsen customer outcomes below this level.
\item The marginal value of new battery capacity is location-dependent, declines with battery penetration, and private versus system-wide benefits can diverge significantly.
\item Stored energy value in existing batteries rises with battery penetration and falls with PV penetration, driven by the local generation-to-storage ratio.
\end{highlights}


\end{frontmatter}

\section{Introduction}
Worldwide electricity demand is increasing rapidly, driven by electrification of end uses such as electric vehicles, heat pumps, and data centers \cite{IEA_Electricity_2025}. Global electricity consumption rose by an estimated 4.3\% in 2024 (vs.\ 2.5\% in 2023) and is projected to grow at about 3.9\% on average through 2025-2027 \cite{IEA_Electricity_2025}. In New England, ISO New England (ISO-NE) attributes most near-term load growth to heat pumps and EVs, with total electricity use projected to increase by about 17\% by 2033 \cite{Reuters2024ISOELeGrowth,ISO-NE2024CELTOverview}. ISO-NE further notes that residential and commercial heat pumps, together with EV charging, are shifting the system toward a more winter-peaking profile \cite{ISO-NE2024CELTOverview,ISO-NE2024Press}. As data-center electricity demand grows rapidly, some recent U.S. state-level proposals would require large data-center developers to support or finance distributed resources in order to offset part of their additional grid burden \cite{Powell2026LinkedIn}.

Accompanying the growth in electricity demand is an increasing role for distributed energy resources (DERs). At the distribution level, rooftop PV, batteries, and other flexible demand are increasingly present. Behind-the-meter solar alone reached roughly 47.7~GW in the U.S.\ by the end of 2023, with about 7.9~GW added in 2023 \cite{IEA-PVPS-USA-2023}. On the utility-scale side, solar has been the largest source of new U.S.\ capacity additions, accompanied by rapid growth in grid-scale batteries \cite{Reuters2025CapacityMix}. While these resources can support decarbonization and flexibility, they also make distribution-system operation more complex. Net demand becomes more volatile, local congestion and voltage issues become more likely, and reliability risks increase when high demand coincides with low renewable output. NERC has therefore identified elevated risk of shortfalls during extreme weather, particularly under such stressed conditions \cite{NERC2024SRA}.

In addition to higher demand and rising distributed energy resource (DER) penetration, forecast uncertainty creates a further operational challenge. In DER-rich feeders, especially those with high PV penetration, actual generation can deviate significantly from day-ahead or hour-ahead expectations because of changing weather conditions. This in turn can create net-load ramps, lead to inefficient battery dispatch, and worsen local congestion, voltage problems, or lost load if the system is not operated adaptively.

Within DER-rich distribution grids, storage is particularly important because it can absorb surplus PV generation, shift energy to peak hours, smooth net-load ramps, and relieve local congestion on stressed feeders \cite{Lamont2013StorageValue,Sisternes2016StorageDecarb}. It can also improve reliability, because storage located close to demand can continue serving load when bulk supply is disrupted, and its value increases when reliability benefits are taken into account \cite{Engelman2025LMVOutage}. At the same time, the value of storage is not uniform across the network. Prior work shows that its locational marginal value depends on net-demand variability, nodal price movements, and network constraints, and that the largest marginal gains are typically obtained at small initial storage capacities \cite{Bose2015LMVStorage}. Storage valuation in distribution grids therefore needs to account for both where storage is installed and how much is already deployed, especially because marginal value tends to decline as penetration increases \cite{Lamont2013StorageValue,Sisternes2016StorageDecarb}.

The operational challenges created by load growth, DER integration, and forecast uncertainty also have important economic consequences. Utilities are responding to load growth, DER interconnections, and resilience needs with record planned capital expenditures for transmission and distribution infrastructure over the second half of the 2020s. Industry capex in recent years has been on the order of \$150--\$180~billion per year in the U.S. \cite{UtilityDive2025Capex,EEI2024CapexPage}. While some grid expansion is unavoidable, more efficient use of existing network capacity can reduce or defer costly overbuilding. Pricing is one important lever. As summarized by \cite{Hinchberger2024CEEPR,CEEPR2022DynamicTariffs}, moving from flat to time-varying tariffs can recover a substantial share of the efficiency lost under flat rates, with estimated savings on the order of \$2~billion per year in avoided deadweight loss under U.S.\ conditions.

One way to address these economic and operational challenges is through local electricity markets (LEMs), as a distribution-level coordination mechanism for DER-rich systems. In a LEM, prices are determined endogenously from local supply, demand, and feeder constraints rather than imposed exogenously through fixed retail tariffs. Using distribution locational marginal prices (d-LMPs), LEMs can reflect both the time- and location-specific value of injections, withdrawals, and flexibility. d-LMPs extend wholesale locational marginal pricing to distribution networks by embedding real losses, voltage limits, and branch congestion directly into nodal prices. This is enabled via an OPF-based market process, settling generation and consumption participants at feeder nodes at their local d-LMPs, and enabling net interchange with the upstream system at wholesale prices. This is particularly attractive in DER-rich feeders, where PV generation, EV charging, and battery dispatch can create rapid and spatially uneven swings in net demand.

This paper studies the performance of a local electricity market (LEM) under PV forecast uncertainty, modeled using an Ornstein--Uhlenbeck process, and analyzes how outcomes vary with PV penetration, storage penetration, and storage location. In addition, the paper quantifies the value of storage within the LEM. The LEM is evaluated using metrics related to lost load, system costs, customer costs, and distribution-operator costs. The results show that the LEM performs favorably on these metrics, particularly in reducing lost load and lowering system costs. To benchmark these results, the LEM is compared with two conventional market designs, including those based on a flat tariff (FT) and real-time pricing (RTP).

\section{Related Work} 

Existing work relevant to this paper can be grouped into three topics: (i) local electricity markets (LEMs) and d-LMP-based coordination on distribution networks, (ii) conventional retail pricing and export-compensation regimes used as benchmark alternatives to LEMs, and (iii) uncertainty-aware operation of DER-rich feeders through stochastic, robust, and receding-horizon methods.

\subsection{Local electricity markets (LEMs)}

A number of studies can be found in the literature related to local electricity markets as a distribution-level coordination mechanism for DER-rich systems \cite{Tan2022LMPReview,Wang2023d-LMPVision,Domenech2024d-LMPDecomp,NairAnnaswamy2023,NairLocalRetail2023,Li2014EVCS_d-LMP,Bai2018d-LMPVoltage,Krishna2024LMPBESS}. References \cite{Tan2022LMPReview,Wang2023d-LMPVision} show how LMPs can be extended from wholesale locational marginal pricing to distribution networks by embedding real losses, voltage limits, and branch congestion directly into nodal prices. References \cite{Tan2022LMPReview,Wang2023d-LMPVision,Domenech2024d-LMPDecomp} show how d-LMPs can be computed under AC power flow formulations and tractable relaxations, and how they can be decomposed into economically interpretable components such as energy, losses, congestion, and voltage terms. Of particular note is the work in \cite{NairAnnaswamy2023,haider2021} where a hierarchical retail LEM mechanism with feeder-level clearing and d-LMP settlement is proposed, and demonstrated on a modified IEEE-123 feeder that coordinated dispatch of PV, batteries, and flexible demand can occur, at reduced operational costs while meeting voltage constraints. References \cite{Li2014EVCS_d-LMP}, \cite{Potter2023ReactivePowerMarket}, and \cite{Krishna2024LMPBESS} have addressed how d-LMP signals can steer EV charging away from congested locations, provide reactive-power and voltage support from DERs, and shape battery dispatch to accommodate network constraints and nodal price patterns, respectively. Despite this progress, several gaps remain. First, although LEMs and d-LMP-based coordination have been studied extensively, relatively little work evaluates their performance under explicitly modeled renewable forecast uncertainty in a way that links forecast errors to realized feeder operation, reliability outcomes, and market results. Second, while prior studies show that d-LMPs influence battery dispatch and that storage can provide local system value, few works quantify the value of storage within a LEM from both a system perspective and a private participant perspective. Very few papers have examined how uncertainty, nodal conditions, and market design together affect operator procurement costs, private cashflows, or the resulting incentives for battery siting and operation.

\subsection{Retail Tariffs} 

Conventional retail tariff structures can be grouped as flat tariffs, time-of-use (TOU) tariffs, critical-peak pricing (CPP), and real-time pricing (RTP) \cite{haider2021}. Of these, flat tariffs apply a uniform retail price over time and do not capture local variations in time; RTP provides a stronger temporal signal as it captures time-varying conditions more directly \cite{haider2021}. Empirical work shows that moving away from flat pricing toward more dynamic pricing can improve efficiency \cite{Hinchberger2024CEEPR}, \cite{FaruquiSergici2010}. Recent evidence suggests that RTP can capture a large share of the efficiency gains from fully dynamic pricing, while TOU and CPP can provide useful intermediate improvements when bill stability and predictability matter \cite{CEEPR2022DynamicTariffs}, \cite{Hinchberger2024CEEPR}. Even though TOU, CPP, and RTP improve signals about when electricity is scarce, they generally do not reflect where distribution-level constraints arise and therefore remain incomplete substitutes for d-LMP-based LEMs. In addition, DER compensation under these signals depends on the export rule, such as net energy metering (NEM) and net energy billing (NEB), which may provide credits at or below the retail rate \cite{CPUC_NEM_NBT}, \cite{NREL_DG_Comp}.

\subsection{Robust feeder operation} 

Uncertainty-aware feeder operation has become increasingly important with increased DER penetration. d-LMP- and OPF-based formulations have been extended using stochastic and robust optimization to capture uncertainty in demand and renewable generation, trading off expected efficiency against conservatism and ex-post feasibility \cite{Zhao2018d-LMPUncertainty},\cite{Robustd-LMP2022}. MPC and receding-horizon methods have been applied to feeder-level OPF and DER coordination as improved forecasts become available leading to lower operating costs and fewer violations \cite{MPCMILP2022,SocialWelfareMPC2020}. Of particular note is the use of mean-reverting stochastic processes, such as Ornstein-Uhlenbeck models, in obtaining tighter forecasts \cite{CarteaFigueroa2005OU} and will be used in this paper.

\subsection{Our contributions}

This paper proposes a d-LMP-settled local electricity market with varied DER penetration including PV generation and batteries. The PV generation is assumed to be stochastic with an explicit PV forecast carried out using a receding-horizon operation. The paper also considers a detailed examination of siting and sizing of batteries and quantifies the value of battery storage in the LEM. A modified IEEE-123 feeder with various DER penetration is studied, and how forecast errors affect reliability, dispatch, and market outcomes is demonstrated. The benefits of LEM are quantified using various metrics including system-level costs, operator costs, and private costs for a range of DER penetration levels. We also benchmark the performance of the LEM against two conventional alternatives, (i) FT with NEM, and (ii) RTP with NEB, using the IEEE-123 feeder for various DER portfolios, full AC feasibility, and a shared realized uncertainty trajectory. We perform extensive sensitivity analyses to understand the cost performance of different markets for various solar and battery penetrations. Finally, we develop and validate two distinct storage valuation approaches using our LEM -- (i) to estimate the marginal value of adding new battery capacity, and (ii) to understand the value of existing storage capacity in the network.

\section{Methodology}\label{sec:methodology}

This section presents the overall LEM market design, uncertainty forecast, and battery sizing and siting. It first introduces the feeder-level formulation of the local electricity market (LEM). It then explains how PV forecast uncertainty is modeled using an Ornstein-Uhlenbeck process and embedded into a two-stage rolling-horizon simulation, in which Stage~1 determines battery schedules based on forecasts and Stage~2 evaluates the realized network-feasible operation under actual PV conditions. This section also includes various cost to assess the performance of the LEM including system cost, customer cost, operator cost, and value of storage.

\subsection{A Local Electricity Market (LEM)}

The LEM we propose is a feeder-level market cleared directly on the primary distribution network. DERs such as PV and battery systems participate through their injections and withdrawals at their respective connection buses, and market clearing is performed using an OPF-based optimization that explicitly enforces distribution-network constraints. These constraints include nodal power balance, line-flow limits, and voltage feasibility, so that the resulting dispatch is not only economically efficient but also physically feasible on the feeder. Overall, the feeder structure can be viewed as a simplified version of our earlier work in \cite{NairAnnaswamy2023}. The market outcome is determined through nodal prices derived from the OPF, interpreted as distribution locational marginal prices (d-LMPs). These prices reflect the time- and location-specific marginal value of electricity given local losses, congestion, and voltage constraints, and are used to settle injections and withdrawals at each bus. In this way, the LEM provides a coordination mechanism that aligns DER operation with feeder conditions rather than relying on exogenous retail tariffs. Thus, such a formulation isolates the core mechanism of interest in this paper: network-feasible, d-LMP-based coordination of storage and PV on a constrained distribution feeder. It also provides a transparent framework for comparing the LEM against alternative tariff-based market designs on the same feeder and under the same uncertainty realization.

\subsection{Modeling PV forecast uncertainty with an Ornstein--Uhlenbeck Process}\label{sec:two_stage}

A central challenge in DER-rich distribution systems is that PV generation is uncertain when operational decisions are made. As realized PV output may differ because of changing weather conditions one needs to carefully carry out PV forecasts. These forecasts are temporally correlated as weather-driven deviations tend to fade rather than exhibit truly random variations. To capture this behavior, this paper models PV forecast error using an Ornstein--Uhlenbeck (OU) process \cite{Zhao2018d-LMPUncertainty}. The OU process is a classical mean-reverting stochastic process that generates forecasts that are serially correlated but revert toward a baseline profile over time. In contrast to independent noise or a random walk, it prevents deviations from drifting permanently away from the underlying PV trajectory. This makes it useful for short-run PV forecasting in a rolling-horizon setting. The OU process is described by
\begin{equation}
    \varepsilon_t = \theta \, \varepsilon_{t-1} + \sigma \, \xi_t,
    \quad \xi_t \sim \mathcal{N}(0,1),
\end{equation}
where \(\varepsilon_t\) is the forecast deviation at time step \(t\), \(\theta \in (0,1)\) controls the persistence of deviations, \(\sigma > 0\) controls the size of new shocks, and \(\xi_t\) is an independent standard Gaussian random variable. Let \(\pi^{\mathrm{real}}_t \in [0,1]\) denote the normalized PV availability profile. A value of \(\hat{\pi}_t=0\) means that no PV generation is available, \(\hat{\pi}_t=1\) means that the full installed PV capacity is available. In this paper, \(\pi^{\mathrm{real}}_t\) serves as the underlying deterministic baseline trajectory. The OU-generated deviation \(\varepsilon_t\) is then used to construct the forecast PV availability \(\hat{\pi}_t\) used in Stage~1:
\begin{equation}
\hat{\pi}_t = \pi^{\mathrm{real}}_t (1 + \varepsilon_t),
\qquad 0 \le \hat{\pi}_t \le 1.
\label{OU_process}
\end{equation}
Thus, the forecast \(\hat{\pi}_t\) is obtained by perturbing the realized baseline profile \(\pi^{\mathrm{real}}_t\) with OU forecast error, while keeping PV availability within the physically meaningful range \(0 \leq \hat{\pi}_t \leq 1\).

\subsection{Market Dispatch}\label{sec:pv_bs_modeling}

To capture the interaction between uncertain PV generation and battery operation, the market is modeled as a two-stage process. Let
\begin{equation}
\mathcal{T}=\{1,2,\dots,T\}
\label{full_modelling_horizon}
\end{equation}
denote the full simulation horizon, where each time step represents 15 minutes. Thus, \(T=96\) corresponds to a 24-hour period. At each decision time \(\tau \in \mathcal{T}\), taken at the beginning of the time step, Stage~1 solves a planning problem over the rolling horizon
\begin{equation}
\mathcal{T}(\tau)=\{\tau,\tau+1,\dots,\tau+H-1\},
\label{forecast_horizon}
\end{equation}
where \(H=32\), corresponding to a planning horizon of 8 hours. 

\subsubsection*{Stage 1: LEM planning problem}

The Stage-1 optimization minimizes wholesale procurement cost, export revenue, and lost-load penalties:
\begin{equation}\label{eq:lem_obj_stage1_short}
\min \sum_{t\in \mathcal{T}(\tau)} \Big(
\lambda^{\mathrm{ISO}}_t P^{\mathrm{imp}}_{t}
-\lambda^{\mathrm{ISO}}_t P^{\mathrm{exp}}_{t}
+ M\sum_{i\in\mathcal{B}} P^{\mathrm{LL}}_{i,t}
\Big),
\end{equation}
where \(P^{\mathrm{imp}}_t\) and \(P^{\mathrm{exp}}_t\) denote feeder imports from and exports to the bulk grid, and \(P^{\mathrm{LL}}_{i,t}\) is lost load at bus \(i\). The objective therefore minimizes net wholesale energy cost and strongly penalizes unserved demand with value-of-lost-load (VOLL) $M$. At decision time \(\tau\), the Stage-1 optimization uses a hybrid PV availability trajectory over the planning horizon \(\mathcal{T}(\tau)\). Specifically, the realized PV availability is used for the current time step, while forecast PV availability is used for the remaining future steps:
\begin{equation}
\tilde{\pi}_{t}=
\begin{cases}
\pi^{\mathrm{real}}_t, & t=\tau,\\[1mm]
\hat{\pi}_t, & t\in\{\tau+1,\tau+2,\dots,\tau+H-1\},
\end{cases}
\label{eq:hybrid_pv_input}
\end{equation}
Thus, the first step of the planning window is evaluated using the currently realized PV output. Because the Stage-1 optimization is solved within the 15-minute dispatch interval and requires on average only about 3 minutes, the realized PV availability for the current step is assumed to be known at the time the decision is made. Forecast values are used for all remaining steps in the horizon. The decision variables are nodal injections \(P_{i,t},Q_{i,t}\), PV outputs \(P^{\mathrm{pv}}_{i,t},Q^{\mathrm{pv}}_{i,t}\), battery power \(P^{\mathrm{bs}}_{i,t},Q^{\mathrm{bs}}_{i,t}\), feeder import and export variables \(P^{\mathrm{imp}}_{t},P^{\mathrm{exp}}_{t},Q^{\mathrm{imp}}_{t},Q^{\mathrm{exp}}_{t}\), battery state of charge (SOC) \(\mathrm{SOC}_{i,t}\), branch flows \(P_{ij,t},Q_{ij,t}\), squared branch currents \(\ell_{ij,t}\), squared voltage magnitudes \(v_{i,t}\), and lost load \(P^{\mathrm{LL}}_{i,t},Q^{\mathrm{LL}}_{i,t}\). The optimization is subject to the following constraints:

\noindent \underline{(i) Nodal power balance:}
For each bus \(i\) and time \(t\in \mathcal{T}(\tau)\),
\begin{align}
P_{i,t} &= P^{\mathrm{pv}}_{i,t} - P^{\mathrm{bs}}_{i,t} - \bigl(L^P_{i,t}-P^{\mathrm{LL}}_{i,t}\bigr), \label{eq:Pbal_stage1_short}\\
Q_{i,t} &= Q^{\mathrm{pv}}_{i,t} - Q^{\mathrm{bs}}_{i,t} - \bigl(L^Q_{i,t}-Q^{\mathrm{LL}}_{i,t}\bigr), \label{eq:Qbal_stage1_short}\\
Q^{\mathrm{LL}}_{i,t} &= \rho_{i,t}\,P^{\mathrm{LL}}_{i,t}, \label{eq:LLratio_stage1_short}
\end{align}
where \(L^P_{i,t}\) and \(L^Q_{i,t}\) denote the active and reactive load at bus \(i\) and time \(t\), respectively, and \(\rho_{i,t}=L^Q_{i,t}/L^P_{i,t}\) if \(L^P_{i,t}>0\), and \(\rho_{i,t}=0\) otherwise.

\noindent \underline{(ii) PV limits:}
For each PV bus \(i\), PV generation may be curtailed.
\begin{align}
0 \le P^{\mathrm{pv}}_{i,t} \le p^{\mathrm{pv}}_{i}\,\tilde{\pi}_{t},\label{eq:pv_cap_stage1_short}\\
-\,q^{\mathrm{pf}}\, P^{\mathrm{pv}}_{i,t} \le Q^{\mathrm{pv}}_{i,t} &\le q^{\mathrm{pf}}\, P^{\mathrm{pv}}_{i,t}, \label{eq:pv_pf_stage1_short}
\end{align}
where \(p^{\mathrm{pv}}_i\) is the installed PV capacity at bus \(i\), and \(q^{\mathrm{pf}}=\tan(\arccos(\mathrm{pf}^{\min}))\) defines the power-factor limit.

\noindent \underline{(iii) Battery limits and SOC dynamics:}
For each battery bus \(i\),
\begin{align}
P^{\mathrm{abs}}_{i,t} &\ge P^{\mathrm{bs}}_{i,t}, \nonumber\\
P^{\mathrm{abs}}_{i,t} &\ge -P^{\mathrm{bs}}_{i,t}, \nonumber\\
P^{\mathrm{abs}}_{i,t} &\le p^{\mathrm{bs}}_{i}, \label{eq:bs_abs_stage1_short}\\
-q^{\mathrm{pf}} P^{\mathrm{abs}}_{i,t}
&\le Q^{\mathrm{bs}}_{i,t}
\le q^{\mathrm{pf}} P^{\mathrm{abs}}_{i,t}, \label{eq:bs_pf_stage1_short}\\
\mathrm{SOC}_{i,t+1}
&=
(1-\delta_{\mathrm{bs}})\,\mathrm{SOC}_{i,t}
+\frac{\eta_{\mathrm{bs}}\Delta t}{C_i}\,P^{\mathrm{bs}}_{i,t}, \label{eq:soc_stage1_short}\\
\mathrm{SOC}^{\min}
&\le \mathrm{SOC}_{i,t}
\le \mathrm{SOC}^{\max}. \label{eq:soc_bounds_stage1_short}
\end{align}
where \(p^{\mathrm{bs}}_i\) is the battery power rating, \(C_i\) is the battery energy capacity, \(\eta_{\mathrm{bs}}\) is battery efficiency, and \(\delta_{\mathrm{bs}}\) is the self-discharge factor. At the beginning of each horizon, the battery state is initialized with the measured current value that is determined in Stage~2. Constraints \cref{eq:bs_abs_stage1_short} are used to keep the battery power-factor constraints convex.

\noindent \underline{(iv) Network constraints:}
The feeder is modeled using the DistFlow equations. For each branch \((i,j)\), \(P_{ij,t}\) and \(Q_{ij,t}\) denote the active and reactive power flow on branch \((i,j)\), \(R_{ij}\) and \(X_{ij}\) denote the branch resistance and reactance, \(v_{i,t}\) is the squared voltage magnitude at bus \(i\), and \(\ell_{ij,t}\) is the squared current magnitude on branch \((i,j)\). The DistFlow equations are
\begin{align}
v_{j,t} &= v_{i,t} - 2\,(R_{ij} P_{ij,t} + X_{ij} Q_{ij,t})
+ (R_{ij}^2+X_{ij}^2)\,\ell_{ij,t}, \label{eq:ohm_stage1_short}\\
P_{j,t} &= -P_{ij,t} + R_{ij}\,\ell_{ij,t} + \sum_{(j,k)\in\mathcal{E}} P_{jk,t}, \label{eq:downstreamP_stage1_short}\\
Q_{j,t} &= -Q_{ij,t} + X_{ij}\,\ell_{ij,t} + \sum_{(j,k)\in\mathcal{E}} Q_{jk,t}, \label{eq:downstreamQ_stage1_short}
\end{align}
where \(\mathcal{E}\) denotes the set of feeder branches. These equations describe voltage drops and downstream active and reactive power balances in the radial network.

\noindent \underline{(v) Feeder import/export split and limits:}
\begin{align}
P_{1,t} &= P^{\mathrm{imp}}_t - P^{\mathrm{exp}}_t, \\
Q_{1,t} &= Q^{\mathrm{imp}}_t - Q^{\mathrm{exp}}_t, \\
0 \le P^{\mathrm{imp}}_t,\quad &0 \le P^{\mathrm{exp}}_t, \\
0 \le Q^{\mathrm{imp}}_t,\quad &0 \le Q^{\mathrm{exp}}_t,
\end{align}
together with feeder import/export bounds.

\noindent \underline{(vi) SOCP relaxation and branch limits:}
\begin{align}
\|(P_{ij,t},Q_{ij,t})\|_2^2 &\le v_{i,t}\,\ell_{ij,t}, \label{eq:rsoc_stage1_short}\\
\|(P_{ij,t},Q_{ij,t})\|_2 &\le S^{\max}_{ij},\qquad
0\le \ell_{ij,t}\le (I^{\max}_{ij})^2, \label{eq:branchlimits_stage1_short}
\end{align}
where \(S^{\max}_{ij}\) is the apparent-power limit and \(I^{\max}_{ij}\) is the current limit of branch \((i,j)\), and with voltage bounds
\begin{equation}
(v^{\min})^2 \le v_{i,t}\le (v^{\max})^2.
\label{eq:voltage_stage1_short}
\end{equation}

\noindent Since all constraints \crefrange{eq:Pbal_stage1_short}{eq:voltage_stage1_short} are affine or second-order cones, the Stage-1 planning problem is a second-order conic program (SOCP) convex relaxation. Overall, the LEM planning problem consists of solving the minimization problem in \cref{eq:lem_obj_stage1_short} subject to the equality constraints in \crefrange{eq:Pbal_stage1_short}{eq:LLratio_stage1_short} and the inequality constraints in \crefrange{eq:pv_cap_stage1_short}{eq:voltage_stage1_short}, which returns as results:

{\footnotesize
\begin{equation}
\left\{
\begin{aligned}
&P^*_{i,t},\;
Q^*_{i,t},\;
P^{\mathrm{pv},*}_{i,t},\;
Q^{\mathrm{pv},*}_{i,t},\;
P^{\mathrm{bs},*}_{i,t},\;
Q^{\mathrm{bs},*}_{i,t},\;
P^{\mathrm{imp},*}_{t},\;
P^{\mathrm{exp},*}_{t},\;\\
&Q^{\mathrm{imp},*}_{t}, \;
Q^{\mathrm{exp},*}_{t},\;
P^*_{ij,t},\;
Q^*_{ij,t},\;
\ell^*_{ij,t},\;
v^*_{i,t},\;
P^{\mathrm{LL},*}_{i,t},\;
Q^{\mathrm{LL},*}_{i,t}
\end{aligned}
\right\},
\qquad t \in \mathcal{T}(\tau)
\label{setpoints}
\end{equation}
}

We note that Stage 1 allows us to determine, in a coordinated way, the PV injections as well as the battery set points among other variables as in \cref{setpoints}. Such a determination can be carried out for all \(t\in T(\tau)\).

\subsubsection*{Stage 2: Realized network operation}\label{stage2}

In Stage~2, the battery setpoint from Stage~1 at time $\tau$ is fixed as $P^{\mathrm{bs},*}_{i,\tau}$ and the feeder is re-optimized at $\tau$, by replacing the upperbound in the constraint in \cref{eq:pv_cap_stage1_short} from $p^{\mathrm{pv}}_{i}\,\tilde{\pi}_{t}$, to $p^{\mathrm{pv}}_{i}\,{\pi}_t^{real}$, and by fixing the decision variable $P^{\mathrm{bs}}_{i,\tau}$ at $\tau$ as in \cref{eq:implemented_bs_stage1_short}.
\begin{equation}
P^{\mathrm{bs},*}_{i,\tau} := P^{\mathrm{bs}}_{i,\tau},
\qquad i\in\mathcal{B}_{\mathrm{bs}},
\label{eq:implemented_bs_stage1_short}
\end{equation}
Thus, Stage 1 uses $\tilde{\pi}_{t}$, the hybrid PV forecast, to determine $P^{\mathrm{bs},*}_{i,\tau}$, while Stage~2 serves as an implementation and evaluation layer in which the battery active-power setpoint is fixed and the realized PV availability \(\pi^{\mathrm{real}}_{\tau}\) (the same deterministic baseline used in \cref{sec:two_stage}) is used to recalculate the remaining contemporaneous setpoints in \cref{setpoints}. This process is repeated for all $\tau \in \mathcal{T}$.

\subsection{Metrics}\label{sec:ex_post}

In order to evaluate the performance of the proposed LEM {\em ex post}, we consider four cost metrics: \emph{system cost}, \emph{private (node-level) cost}, \emph{customer cost}, and \emph{operator net cost}. All costs are assumed to be measured in \$/step.

\subsubsection{System cost (resource cost)}
The \emph{system cost} includes only real resource costs and excludes internal transfers:
\begin{align}
C^{\mathrm{sys}}_t
&:= \lambda^{\mathrm{ISO}}_t\Big(P^{\mathrm{imp}}_t - P^{\mathrm{exp}}_t\Big)
\;+\;
M\sum_{i\in\mathcal{B}} P^{\mathrm{LL}}_{i,t},
\label{eq:sys_cost_step}\\[1mm]
C^{\mathrm{sys}}
&:=\sum_{t\in\mathcal{T}_{\text{tot}}} C^{\mathrm{sys}}_t .
\label{eq:sys_cost_total}
\end{align}
Here, $\lambda^{\mathrm{ISO}}_t$ is the wholesale electricity price at time $t$, and $M$ is the value-of-lost-load (VOLL) penalty. The first term gives net wholesale procurement cost, and the second term penalizes unserved energy.

\subsubsection{Private (node-level) cost}
For customer outcomes, behind-the-meter imports and exports are computed at each node. Let
\begin{equation}
D_{i,t} := L^P_{i,t}-P^{\mathrm{LL}}_{i,t}\qquad (i\in\mathcal{B})
\end{equation}
denote served demand, where $L^P_{i,t}$ is active demand at bus $i$ and time $t$. Define
\begin{equation}
P^{\mathrm{bs,ch}}_{i,t}:=[P^{\mathrm{bs}}_{i,t}]_+,\qquad
P^{\mathrm{bs,dis}}_{i,t}:=[-P^{\mathrm{bs}}_{i,t}]_+,
\end{equation}
where $[x]_+ := \max\{x,0\}$. Local supply and local demand are then
\begin{equation}
S_{i,t}:=P^{\mathrm{pv}}_{i,t}+P^{\mathrm{bs,dis}}_{i,t},\qquad
K_{i,t}:=D_{i,t}+P^{\mathrm{bs,ch}}_{i,t},
\end{equation}
where $P^{\mathrm{pv}}_{i,t}$ is PV active power. Thus, $S_{i,t}$ is local supply from PV and battery discharge, while $K_{i,t}$ is local demand from served load and battery charging. The node's net meter injection is
\[
p^{\mathrm{net}}_{i,t}:=S_{i,t}-K_{i,t},
\]
with corresponding import and export
\begin{equation}
P^{\mathrm{exp}}_{i,t}:=[p^{\mathrm{net}}_{i,t}]_+,\qquad
P^{\mathrm{imp}}_{i,t}:=[-p^{\mathrm{net}}_{i,t}]_+.
\label{eq:nodal_imp_exp}
\end{equation}

The import price $\lambda^{\mathrm{imp}}_{i,t}$ and export credit $\lambda^{\mathrm{exp}}_{i,t}$ in the LEM correspond to $\lambda^{\mathrm{d-lmp}}_{i,t}$. The \emph{private cost} at node $i$ at each time and over the total duration are defined as
\begin{equation}
C^{\mathrm{priv}}_{i,t}
:= \lambda^{\mathrm{imp}}_{i,t}\,P^{\mathrm{imp}}_{i,t}
-\lambda^{\mathrm{exp}}_{i,t}\,P^{\mathrm{exp}}_{i,t},
\qquad
C^{\mathrm{priv}}_{i}:=\sum_{t\in\mathcal{T}_{\text{tot}}} C^{\mathrm{priv}}_{i,t}.
\label{eq:private_cost_def}
\end{equation}
This is the net electricity bill at node $i$: imports are paid, while exports receive credits. We further decompose grid imports into demand-related and battery-charging-related parts:
\begin{align}
P^{\mathrm{grid\to dem}}_{i,t} &:= [D_{i,t}-S_{i,t}]_+,
\\
P^{\mathrm{grid\to bs}}_{i,t} &:= \Big[P^{\mathrm{bs,ch}}_{i,t}-[S_{i,t}-D_{i,t}]_+\Big]_+,
\end{align}
so that $P^{\mathrm{imp}}_{i,t}=P^{\mathrm{grid\to dem}}_{i,t}+P^{\mathrm{grid\to bs}}_{i,t}$.
Similarly, exports are split into PV and battery components:
\begin{align}
P^{\mathrm{pv\to exp}}_{i,t} &:= [P^{\mathrm{pv}}_{i,t}-K_{i,t}]_+,
\\
P^{\mathrm{bs\to exp}}_{i,t} &:= \Big[P^{\mathrm{bs,dis}}_{i,t}-[K_{i,t}-P^{\mathrm{pv}}_{i,t}]_+\Big]_+,
\end{align}
so that $P^{\mathrm{exp}}_{i,t}=P^{\mathrm{pv\to exp}}_{i,t}+P^{\mathrm{bs\to exp}}_{i,t}$. Using this breakdown, we get the following:
\begin{equation}
\begin{aligned}
C^{\mathrm{priv}}_{i,t}
&=
\underbrace{\lambda^{\mathrm{imp}}_{i,t} P^{\mathrm{grid\to dem}}_{i,t}}_{\text{demand-from-grid}}
+
\underbrace{\lambda^{\mathrm{imp}}_{i,t} P^{\mathrm{grid\to bs}}_{i,t}}_{\text{battery charging}}
\\
&\quad-
\underbrace{\lambda^{\mathrm{exp}}_{i,t} P^{\mathrm{pv\to exp}}_{i,t}}_{\text{PV export credit}}
-
\underbrace{\lambda^{\mathrm{exp}}_{i,t} P^{\mathrm{bs\to exp}}_{i,t}}_{\text{battery export credit}} .
\end{aligned}
\label{eq:private_cost_components}
\end{equation}

\subsubsection{Customer cost (aggregate)}
The \emph{customer cost} is the sum of private costs across all buses:
\begin{equation}
C^{\mathrm{cust}}_t := \sum_{i\in\mathcal{B}} C^{\mathrm{priv}}_{i,t},
\qquad
C^{\mathrm{cust}} := \sum_{t\in\mathcal{T}_{\text{tot}}} C^{\mathrm{cust}}_t
= \sum_{i\in\mathcal{B}} C^{\mathrm{priv}}_{i}.
\label{eq:customer_cost}
\end{equation}

\subsubsection{Operator net cost}
The operator pays the resource cost and receives the opposite of the customer bill through settlement. The \emph{operator net cost} is then
\begin{align}
C^{\mathrm{op}}_t
&:= C^{\mathrm{sys}}_t - C^{\mathrm{cust}}_t,
\label{eq:operator_cost_step}\\[1mm]
C^{\mathrm{op}}
&:= \sum_{t\in\mathcal{T}_{\text{tot}}} C^{\mathrm{op}}_t
= C^{\mathrm{sys}} - C^{\mathrm{cust}}.
\label{eq:operator_cost_total}
\end{align}
With this convention, $C^{\mathrm{op}}>0$ means the operator incurs a net cost, while $C^{\mathrm{op}}<0$ means net profit. Using the methodology above, in the subsequent sections, we evaluate the following:
\begin{itemize}
\item \textbf{Reliability impacts}: Lost load or Expected Energy Not Served (EENS) under varying DER penetrations and PV uncertainty.
\item \textbf{Economic outcomes}: System cost, operator net costs, and customer bills.
\item \textbf{Marginal storage value}: From both the private and system-wide perspectives.
\end{itemize}

\section{Results and Discussion}

\subsection{The Use Case}
\begin{figure}[htbp]
    \centering
    \includegraphics[width=0.5\linewidth]{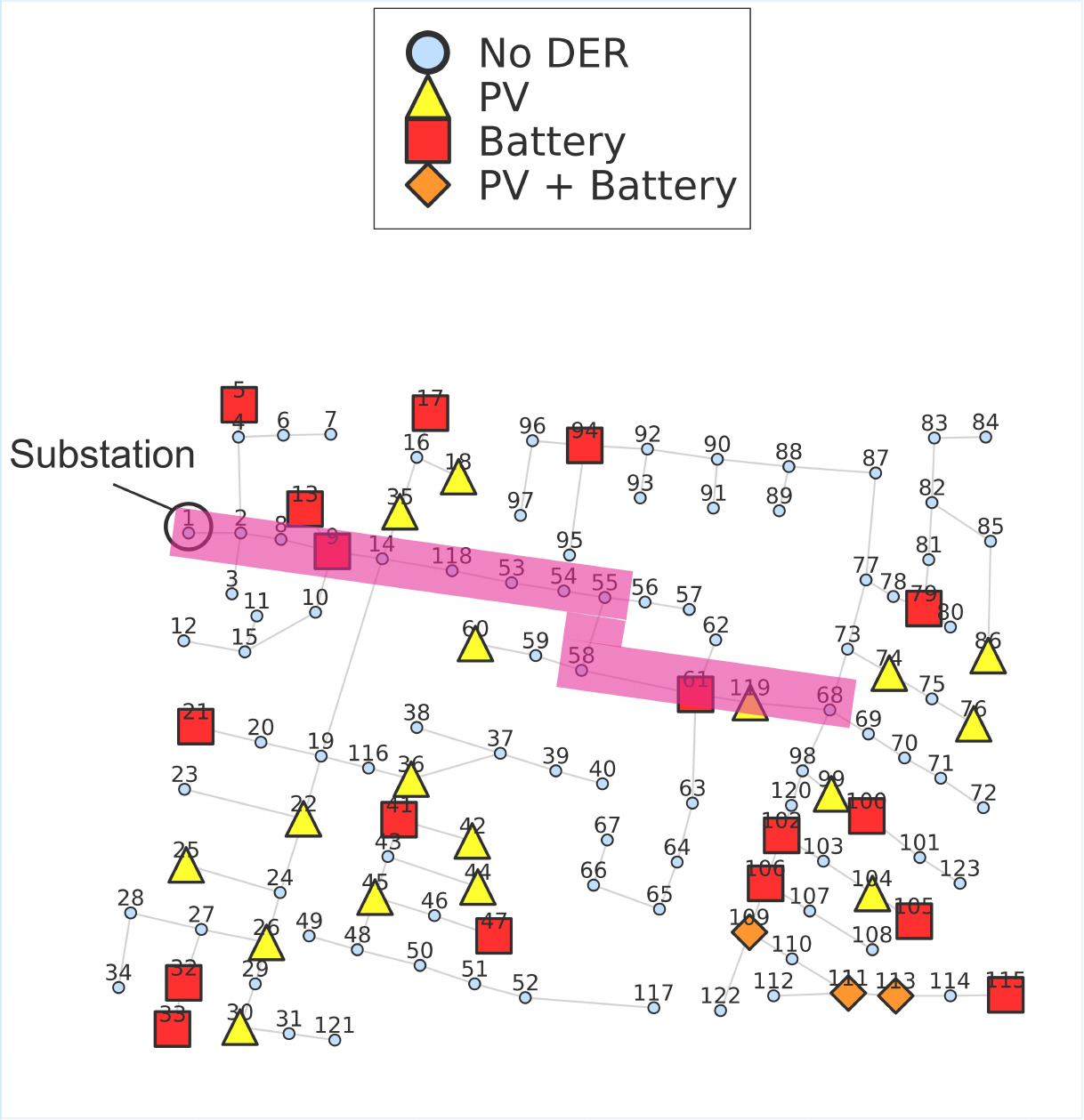}
    \caption{IEEE-123 network with backbone lines highlighted (violet color) and random DER siting for 20 DERs.}
    \label{fig:ieee123_backbone}
\end{figure}

The proposed LEM is evaluated on a modified IEEE-123 feeder. A total of 20 battery storages and 20 PV plants are randomly placed across the feeder, as shown in \cref{fig:ieee123_backbone}. Their sizes are scaled through PV and battery penetration levels, defined relative to the feeder peak load \(P^{\mathrm{peak}}\). Let \(\mathcal{B}_{\mathrm{pv}}\) denote the set of PV buses, \(\mathcal{B}_{\mathrm{bs}}\) the set of battery buses, \(p^{\mathrm{pv}}_i\) the installed PV capacity at bus \(i\), and \(p^{\mathrm{bs}}_i\) the battery power rating at bus \(i\). The penetration levels are defined as
\begin{equation}
\gamma^{\mathrm{pv}} := \frac{\sum_{i\in\mathcal{B}_{\mathrm{pv}}} p^{\mathrm{pv}}_i}{P^{\mathrm{peak}}},
\qquad
\gamma^{\mathrm{bs}} := \frac{\sum_{i\in\mathcal{B}_{\mathrm{bs}}} p^{\mathrm{bs}}_i}{P^{\mathrm{peak}}},
\end{equation}
where \(P^{\mathrm{peak}}=3.68~\mathrm{MW}\) is the feeder peak load. Thus, for example, a PV penetration of \(50\%\) means that the total installed PV capacity equals \(0.5\,P^{\mathrm{peak}}\), and analogously for battery power capacity. For a given penetration level, the total installed PV and battery capacity is distributed equally across the 20 PV plants and 20 batteries, respectively. Loads are present at 85 nodes. All battery systems use the same fixed \(C\)-rate \(c_{\mathrm{bs}}=0.25\), corresponding to 4-hour batteries. The simulation horizon ($\mathcal{T}$) is 24 hours with a timestep of \(\Delta t=15\) minutes, and the rolling forecast horizon is \(H=8\) hours. The wholesale energy prices at the PCC ($\lambda^{\mathrm{ISO}}_t$) were set using 5-minute LMP data from ISO-NE on August 6, 2025, which was downsampled to 15-minute granularity \cite{ISO-NE_FiveMinuteLMPs}. The normalized PV availability profile $\pi^{\mathrm{real}}_t$ was constructed from Boston data for August 6, 2025, a sunny day. This same profile serves as both the deterministic baseline for the OU forecast generation \eqref{OU_process} and as the realized PV input to Stage 2 (\ref{stage2}). VOLL was set to 10,000 \$/MWh, as used by MISO \cite{MISO_Scarcity_VOLL2024}.

The IEEE 123-bus feeder was converted into a balanced single-phase equivalent network. To do this, all switches were assumed to be in their normal positions and all lines were treated as transposed three-phase lines. The line configurations were then converted into symmetric-component form, and only the positive-sequence impedance was retained for each branch. The original phase-specific spot loads were also converted into balanced nodal loads. This made it possible to model the feeder as a single-phase network while preserving the main topology and overall loading characteristics of the original system. Two static thermal classes are assigned to the IEEE-123 feeder to induce realistic congestion at modest DER penetrations: feeder laterals at \(I^{\max}_{ij}=325~\mathrm{A}\) and \emph{backbone} segments at \(I^{\max}_{ij}=750~\mathrm{A}\). These levels fall within typical distribution-system assumptions \cite{NexansACSRData}. Apparent-power limits are computed as
\[
S^{\max}_{ij} \;=\; \sqrt{3}\,V^{\mathrm{nom}}_{\mathrm{LL}}\, I^{\max}_{ij},
\qquad
S^{\max,\mathrm{pu}}_{ij} \;=\; \frac{S^{\max}_{ij}}{S_{\!\text{base}}},
\]
using the feeder nominal line-to-line voltage \(V^{\mathrm{nom}}_{\mathrm{LL}}=4.16~\mathrm{kV}\) and base \(S_{\!\text{base}}=5~\mathrm{MVA}\).
\cref{fig:ieee123_backbone} highlights which branches are designated as ``backbone''. To avoid creating an artificial bottleneck caused by the simplified assignment of only two thermal classes, the current limit of one lateral branch (between nodes 14 and 19) was increased by 10\%. This minor adjustment improves numerical robustness while preserving the qualitative congestion patterns of the feeder. Since the increase is small relative to the backbone-to-lateral rating difference, it does not materially affect the conclusions of the study. These ratings are rather conservative and time-invariant (no weather dependence) to create binding thermal constraints and occasional lost load for low DER penetrations, enabling sensitivity analyses of congestion and d-LMP formation.

\subsubsection{Simulation of the OU process}

\begin{figure}[htbp]
    \centering
    \includegraphics[width=0.85\linewidth]{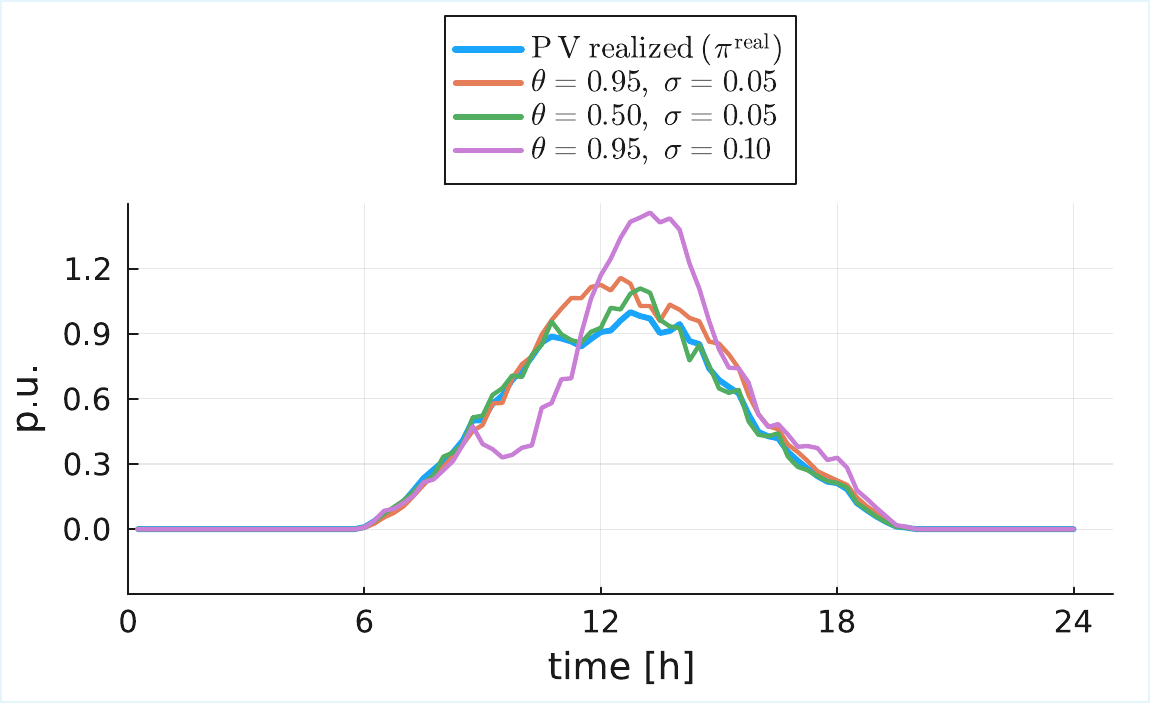}
    \caption{OU parameter variants ($\sigma$, $\theta$).}
    \label{fig:ou_vary_persistence}
\end{figure}

PV inflow curves are generated by applying \cref{OU_process}. To ensure comparability across parameter settings, the same random seed is used in each configuration, so differences across cases are driven only by the OU parameters $\theta$ and $\sigma$. \cref{fig:ou_vary_persistence} illustrates the role of varying persistence $\theta$ as well as $\sigma$. When $\theta$ is low, deviations decay quickly and appear as short-lived fluctuations. When $\theta$ is high, shocks persist and overlap, producing smoother but larger swings around the deterministic PV baseline. Small $\sigma$ produces mild deviations, whereas larger $\sigma$ leads to more pronounced fluctuations because each new shock is larger. This also shows that reducing $\sigma$ as $\theta$ increases can equalize the stationary variance across cases. This helps isolate the effect of persistence from the effect of variance magnitude: $\theta$ controls the correlation structure of deviations, while $\sigma$ sets the scale of newly introduced shocks.
Taken together, these examples provide intuition for PV inflow modeling with an OU process: $\theta$ governs the memory of deviations, while $\sigma$ governs their volatility.

\begin{itemize}
  \item \textbf{$\theta$ (persistence / memory):} proxy for how slowly cloud-driven deviations evolve relative to the timestep $\Delta t$. Values near 1 imply long-lived deviations; lower values imply faster decay.
  \item \textbf{$\sigma$ (volatility / shock size):} proxy for the magnitude of irradiance fluctuations at timestep $\Delta t$. Larger values represent stronger short-run variability; smaller values represent smoother conditions.
\end{itemize}

For all simulations, a persistence $\theta$ value of 0.95 and a Gaussian noise $\sigma$ value of 0.05 were used.

\subsubsection{Battery valuation}\label{BSMV}
The marginal storage value is estimated by perturbing battery energy capacity at a single bus and re-simulating
under identical exogenous trajectories (loads, prices, and OU seed). Fix a bus $i\in\mathcal{B}_{\mathrm{bs}}$ and bump size $\varepsilon>0$ (MWh).
Let the baseline capacity be $C_i$ and the perturbed capacity be
\begin{equation}
C'_i = C_i + \varepsilon.
\end{equation}
To preserve the energy-to-power ratio, the power limits are scaled using a fixed $C$-rate $c_{\mathrm{bs}}$:
\begin{equation}
p^{\mathrm{bs}\,'}_i = p^{\mathrm{bs}}_i + c_{\mathrm{bs}}\varepsilon.
\end{equation}
To avoid injecting ``free energy'', the initial stored energy is kept fixed. Let $C^{\mathrm{sys}}(0)$ and $C^{\mathrm{sys}}(\varepsilon)$ denote total system cost under the baseline and perturbed runs.
The marginal system value (in \$/MWh of added capacity) is estimated by
\begin{equation}
\mathrm{MV}^{\mathrm{sys}}_i \approx \frac{C^{\mathrm{sys}}(0)-C^{\mathrm{sys}}(\varepsilon)}{\varepsilon}.
\end{equation}
Analogously, marginal \emph{customer/private} value replaces $C^{\mathrm{sys}}$ with $C^{\mathrm{cust}}$ or $C^{\mathrm{priv}}_i$.

\subsection{Benchmark models}
To evaluate the performance of the LEM against established retail tariff designs, two benchmark models are introduced: a flat-tariff model and a real-time pricing (RTP) model.

\subsubsection{Flat tariff benchmark (NEM)}\label{sec:flat_benchmark}

The flat benchmark replaces the d-LMP settlement with a uniform retail rate $\phi^{\mathrm{flat}}$ and net energy metering (NEM), i.e., imports and exports are valued at the same rate. The FT is set exogenously to the retail flat tariff used by Eversource \cite{Eversource_RatesTariffs}:
\begin{equation}
\phi^{\mathrm{flat}} = 148.84 \ \text{\$/MWh},
\label{eq:flat_rate}
\end{equation}
which is converted into the model's internal units (\$/p.u.-step) before optimization. In the FT benchmark, batteries are scheduled locally (per battery bus) using a receding-horizon linear program. Network constraints are ignored in Stage~1 and are enforced in Stage~2. At time $\tau$, the planner uses the measured SOC and the PV forecast
$\tilde\pi_t$ over the window $T(\tau)$. Here, $i$ indexes buses, $\tau$ is the current decision time, $T(\tau)=\{\tau,\dots,\tau+H-1\}$ is the planning horizon, and $\tilde\pi_t\in[0,1]$ is the forecast normalized PV availability at time $t$. The (forecast) behind-the-meter net load at bus $i$ is defined as
\begin{equation}
\hat n_{i,t} := L^P_{i,t} - p^{\mathrm{pv}}_{i}\tilde\pi_t,
\qquad t\in T(\tau),
\end{equation}
where $L^P_{i,t}$ is active demand at bus $i$ and time $t$, and $p^{\mathrm{pv}}_{i}$ is the installed PV capacity at bus $i$. For each battery bus $i\in\mathcal{B}_{\mathrm{bs}}$, battery power $P^{\mathrm{bs}}_{i,t}$ and
import/export variables $(u^{+}_{i,t},u^{-}_{i,t})\ge 0$ are chosen such that
\begin{equation}\label{eq:btm_balance_flat}
u^{+}_{i,t}-u^{-}_{i,t}=\hat n_{i,t}+P^{\mathrm{bs}}_{i,t},
\qquad t\in T(\tau),
\end{equation}
where $\mathcal{B}_{\mathrm{bs}}$ is the set of battery buses, $P^{\mathrm{bs}}_{i,t}$ is battery active power, $u^{+}_{i,t}$ is behind-the-meter grid import, and $u^{-}_{i,t}$ is behind-the-meter grid export. The same sign convention as in \cref{sec:pv_bs_modeling} is used, for the sign of $P^{\mathrm{bs}}_{i,t}$ as before. SOC dynamics and bounds are
\begin{align}
\mathrm{SOC}_{i,t+1} &= (1-\delta_{\mathrm{bs}})\,\mathrm{SOC}_{i,t}+\frac{\eta_{\mathrm{bs}}\Delta t}{C_i}P^{\mathrm{bs}}_{i,t} \label{eq:soc_update}\\
\mathrm{SOC}^{\min}\le \mathrm{SOC}_{i,t} &\le \mathrm{SOC}^{\max},\qquad
-p^{\mathrm{bs}}_i\le P^{\mathrm{bs}}_{i,t}\le p^{\mathrm{bs}}_i .
\end{align}
Here, $\mathrm{SOC}_{i,t}$ is the battery state of charge, $\delta_{\mathrm{bs}}$ is the per-step self-discharge factor, $\eta_{\mathrm{bs}}$ is the battery efficiency parameter, $\Delta t$ is the timestep length, $C_i$ is the battery energy capacity at bus $i$, and $p^{\mathrm{bs}}_i$ is the battery power rating. Under NEM, the retail price is the same for imports and exports:
\begin{equation}
\lambda^{\mathrm{imp}}_t=\lambda^{\mathrm{exp}}_t=\phi^{\mathrm{flat}},
\end{equation}
where $\lambda^{\mathrm{imp}}_t$ and $\lambda^{\mathrm{exp}}_t$ are the retail import price and export credit, respectively. The objective minimizes retail energy charges plus a linear throughput degradation proxy:
\begin{equation}\label{eq:flat_stage1_obj}
\min \sum_{t\in T(\tau)}
\Big(\phi^{\mathrm{flat}}\,u^{+}_{i,t}-\phi^{\mathrm{flat}}\,u^{-}_{i,t}
+ c^{\mathrm{deg}}\,|P^{\mathrm{bs}}_{i,t}|\Big).
\end{equation}
Here, $c^{\mathrm{deg}}$ is the marginal battery degradation cost coefficient. Only the first-step setpoint $P^{\mathrm{bs}}_{i,\tau}$ is implemented, and the procedure repeats at $\tau+1$. Stage~2 is identical to that used for the LEM (\cref{stage2}). 
This per-node formulation is inherently egoistic: each battery minimizes its own retail bill independently, with no visibility into network conditions or the dispatch of other batteries.

\subsubsection{Real-Time Pricing (RTP) benchmark (NEB-style export credit)}\label{sec:rtp_benchmark}

The RTP benchmark uses a deterministic retail import price path derived from the temporal profile of the PCC LMP, but rescaled so that its time average equals the flat retail tariff. Let $\lambda^{\mathrm{ISO}}_t$ denote the PCC wholesale price profile. The RTP import tariff is defined as
\begin{equation}
\lambda^{\mathrm{rt}}_t
=
\phi^{\mathrm{flat}}
\frac{\lambda^{\mathrm{ISO}}_t}
{\frac{1}{|\mathcal{T}|}\sum_{s\in\mathcal{T}}\lambda^{\mathrm{ISO}}_s},
\qquad t\in\mathcal{T},
\label{eq:rtp_scaled}
\end{equation}
so that
\begin{equation}
\frac{1}{|\mathcal{T}|}\sum_{t\in\mathcal{T}} \lambda^{\mathrm{rt}}_t
=
\phi^{\mathrm{flat}}.
\end{equation}
Here, $\lambda^{\mathrm{rt}}_t$ is the RTP retail import price at time $t$, obtained by rescaling the PCC LMP profile so that its mean equals the FT, and $\beta$ is the export-credit factor. In the implementation, $\beta = 0.25$. The same per-battery receding-horizon convex optimization problem used in the flat benchmark is solved, but now with time-varying retail import prices and a discounted export credit. At time $\tau$, the forecast net load $\hat n_{i,t}$ is defined as in \cref{eq:btm_balance_flat}. For each $i\in\mathcal{B}_{\mathrm{bs}}$, the Stage-1 optimization is
\begin{align}
\min\;& \sum_{t\in T(\tau)}
\Big(\lambda^{\mathrm{rt}}_{t}\,u^{+}_{i,t}-\lambda^{\mathrm{exp}}_{t}\,u^{-}_{i,t}
+ c^{\mathrm{deg}}\,|P^{\mathrm{bs}}_{i,t}|\Big)\label{eq:rtp_stage1_obj}\\
\text{s.t. }&
u^{+}_{i,t}-u^{-}_{i,t}=\hat n_{i,t}+P^{\mathrm{bs}}_{i,t},\quad u^{+}_{i,t},u^{-}_{i,t}\ge 0,\\
&\text{SOC dynamics and bounds as in \cref{eq:soc_stage1_short,eq:soc_bounds_stage1_short}}.\nonumber
\end{align}
In this formulation, $u^{+}_{i,t}$ and $u^{-}_{i,t}$ are again the behind-the-meter import and export variables, $P^{\mathrm{bs}}_{i,t}$ is battery power, and $c^{\mathrm{deg}}$ is the degradation-cost coefficient. Prices satisfy:
\begin{equation}
\lambda^{\mathrm{imp}}_t=\lambda^{\mathrm{rt}}_t,\qquad
\lambda^{\mathrm{exp}}_t=\beta\,\lambda^{\mathrm{rt}}_t,\qquad \alpha\in(0,1].
\end{equation}
Thus, imports are charged at the RTP retail price, while exports receive only a fraction $\beta$ of that RTP price. Stage~2 is identical to that used for the LEM (\cref{stage2}).
As in the FT case, this formulation is egoistic: each battery responds solely to local price signals, with no network awareness or inter-node coordination.

\subsubsection{Settlement prices}\label{sec:settlement_prices}

Three market and tariff designs are compared that differ only in (i) the prices used for settlement and (ii) the Stage-1 battery scheduling logic (\cref{tab:market-designs}). In all cases, settlement is computed \emph{ex post} from the realized quantities produced by Stage~2: PCC import/export $(P^{\mathrm{imp}}_\tau,P^{\mathrm{exp}}_\tau)$ and realized nodal injections. In the LEM, nodal settlement prices are taken to be the distribution locational marginal prices (d-LMPs) implied by the OPF. Let the active nodal power-balance constraints be $h^P_{i,t}(x)=0$ with dual multiplier $\lambda^P_{i,t}$. The d-LMP for real power at bus $i$ and time $t$ is given by the active-power multiplier $\lambda^P_{i,t}$. Economically, $\lambda^P_{i,t}$ is interpreted as the marginal change in the OPF objective induced by a marginal increase in real-power injection at bus $i$ (or equivalently, by a marginal decrease in withdrawal, depending on sign conventions) \cite{BoydVandenberghe2004}. Using the Stage~2 realized quantities, settlement is carried out as follows:
\begin{itemize}
\item LEM: nodal revenues/costs are computed using d-LMPs $\lambda^{\mathrm{d\text{-}LMP}}_{i,\tau}$.
\item Flat (NEM): the retail charge is computed using $\phi^{\mathrm{flat}}$ for both imports and exports.
\item RTP (NEB): the retail import price is set to $\lambda^{\mathrm{rt}}_\tau$ and the export credit is set to $\beta\lambda^{\mathrm{rt}}_\tau$.
\end{itemize}

All price settlement differences are summarized in \cref{tab:market-designs}.

\begin{table*}[t]
\centering
\caption{Comparison of the three market designs. Each mode differs only in Stage-1 decision logic and settlement prices.}
\label{tab:market-designs}
\small
\begin{tabular}{p{1.0cm}p{3.2cm}p{3.0cm}p{6.0cm}}
\toprule
\textbf{Mode} & \textbf{Import price} & \textbf{Export credit} & \textbf{Stage-1 decision} \\
\midrule
LEM &
Nodal d-LMP settlement &
Nodal d-LMP settlement &
Centralized rolling-horizon AC-OPF; network-aware co-optimization of all DERs \\
FT &
Constant flat retail tariff $\phi^{\mathrm{flat}} = 148.84$ \$/MWh &
Same FT $\phi^{\mathrm{flat}}$ (NEM) &
Decentralized per-battery receding-horizon LP; egoistic, no network constraints or inter-node coordination \\
RTP &
Scaled RTP import price $\lambda_t^{\mathrm{rt}} \propto \lambda_t^{\mathrm{ISO}}$, normalized so that $\frac{1}{|\mathcal{T}|}\sum_t \lambda_t^{\mathrm{rt}}=\phi^{\mathrm{flat}}$ &
$\beta\,\lambda_t^{\mathrm{rt}}$ (NEB) &
Decentralized per-battery receding-horizon LP; egoistic, no network constraints or inter-node coordination \\
\bottomrule
\end{tabular}
\end{table*}

\subsection{Results}

Using the metrics defined in \cref{sec:ex_post}, the effect of PV forecast uncertainty on the LEM is assessed and the LEM is compared with the two benchmark designs, FT + NEM and RTP + NEB, on the same IEEE-123 feeder under identical DER portfolios and the same realized PV uncertainty trajectory. The results are organized around three questions: (i) how market design affects reliability through lost load; (ii) how costs are distributed across the operator, customers, and the system as a whole; and (iii) how the private and system value of additional battery capacity in the LEM depend on location and baseline battery penetration. The results are presented as monthly bills, which are obtained by scaling the one-day simulation results to a 30-day month (standard deviations of days assumed as independent events and scaled by multiplying with $\sqrt{30}$). Means, standard deviations, and confidence intervals are computed across 10 simulations with different OU realizations of PV forecast error. Since the input data correspond to a sunny summer day in Boston, the resulting monthly costs should be interpreted as representative of a good-weather summer month rather than an annual average.

\subsubsection{Cost outcome over a range of PV realizations and under different market designs}

\begin{figure*}[htbp]
  \centering
  \begin{subfigure}[t]{0.32\linewidth}
    \centering
    \includegraphics[width=\linewidth]{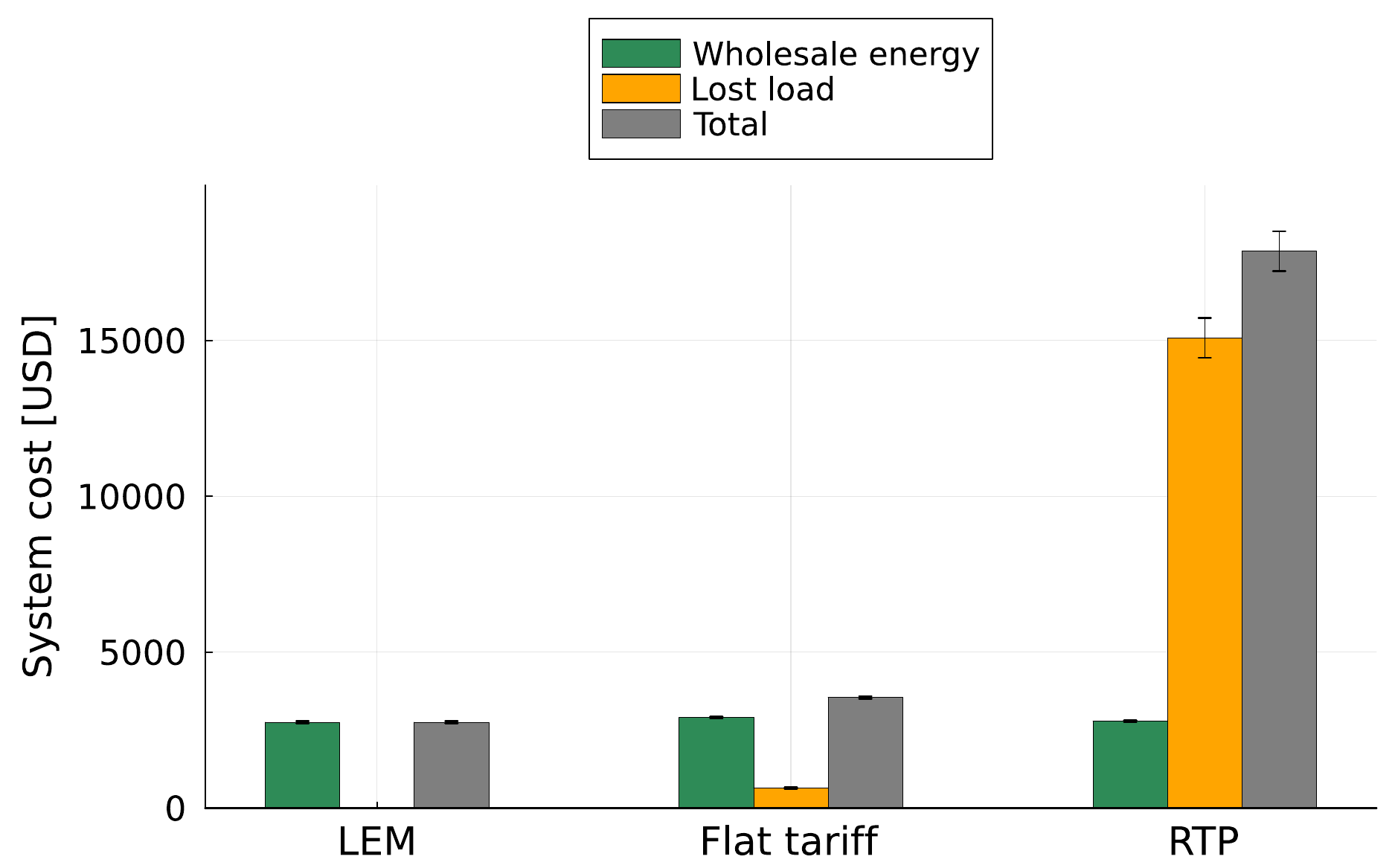}
    \caption{Daily system cost. LEM has 22\% lower system cost compared to FT, and 85\% lower than RTP.}
    \label{fig:systemcost_50_50}
  \end{subfigure}
  \hfill
  \begin{subfigure}[t]{0.32\linewidth}
    \centering
    \includegraphics[width=\linewidth]{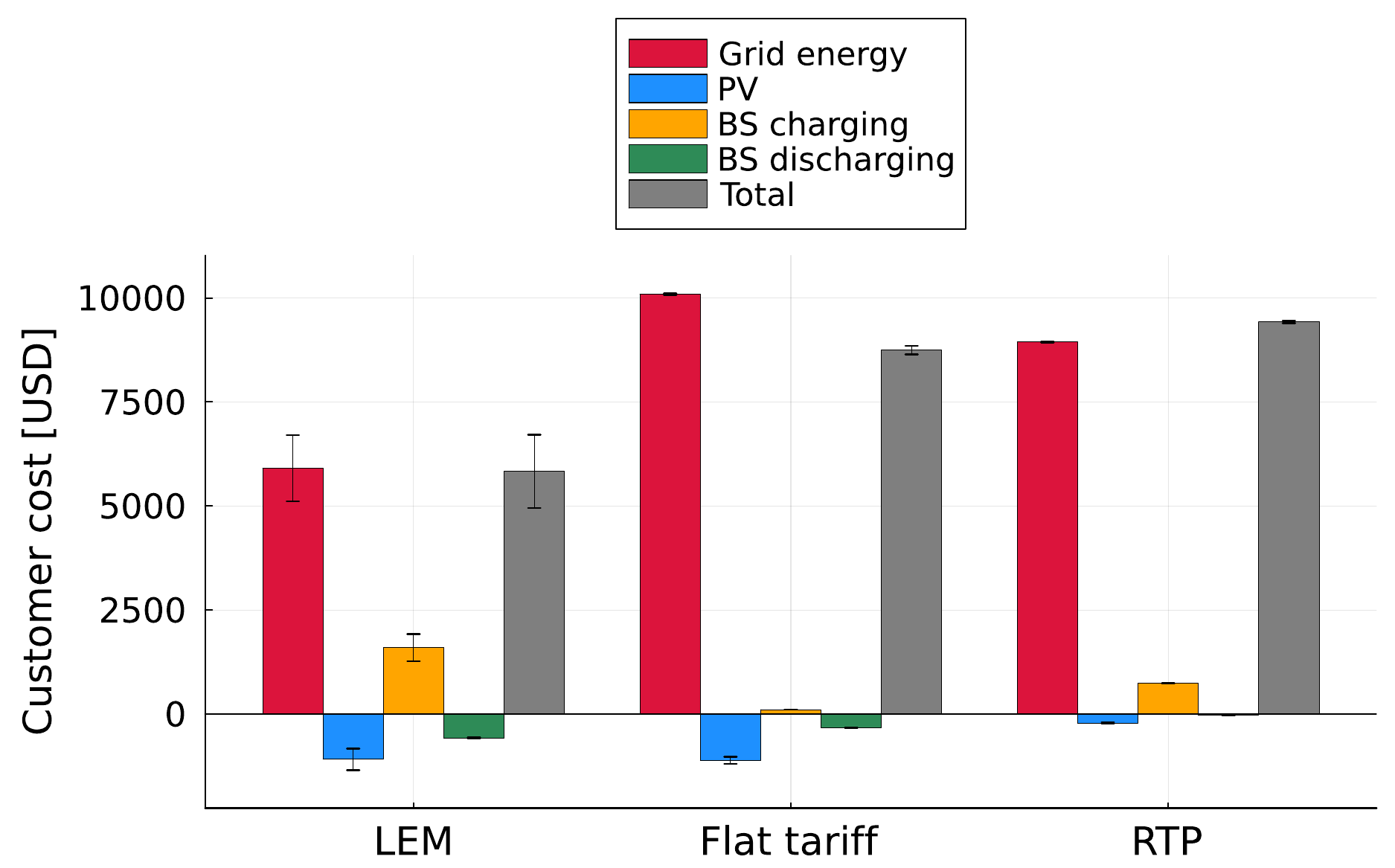}
    \caption{Daily customer cost components. LEM has 33\% lower customer cost compared to FT, and 38\% lower than RTP.}
    \label{fig:customercost_50_50}
  \end{subfigure}
  \hfill
  \begin{subfigure}[t]{0.32\linewidth}
    \centering
    \includegraphics[width=\linewidth]{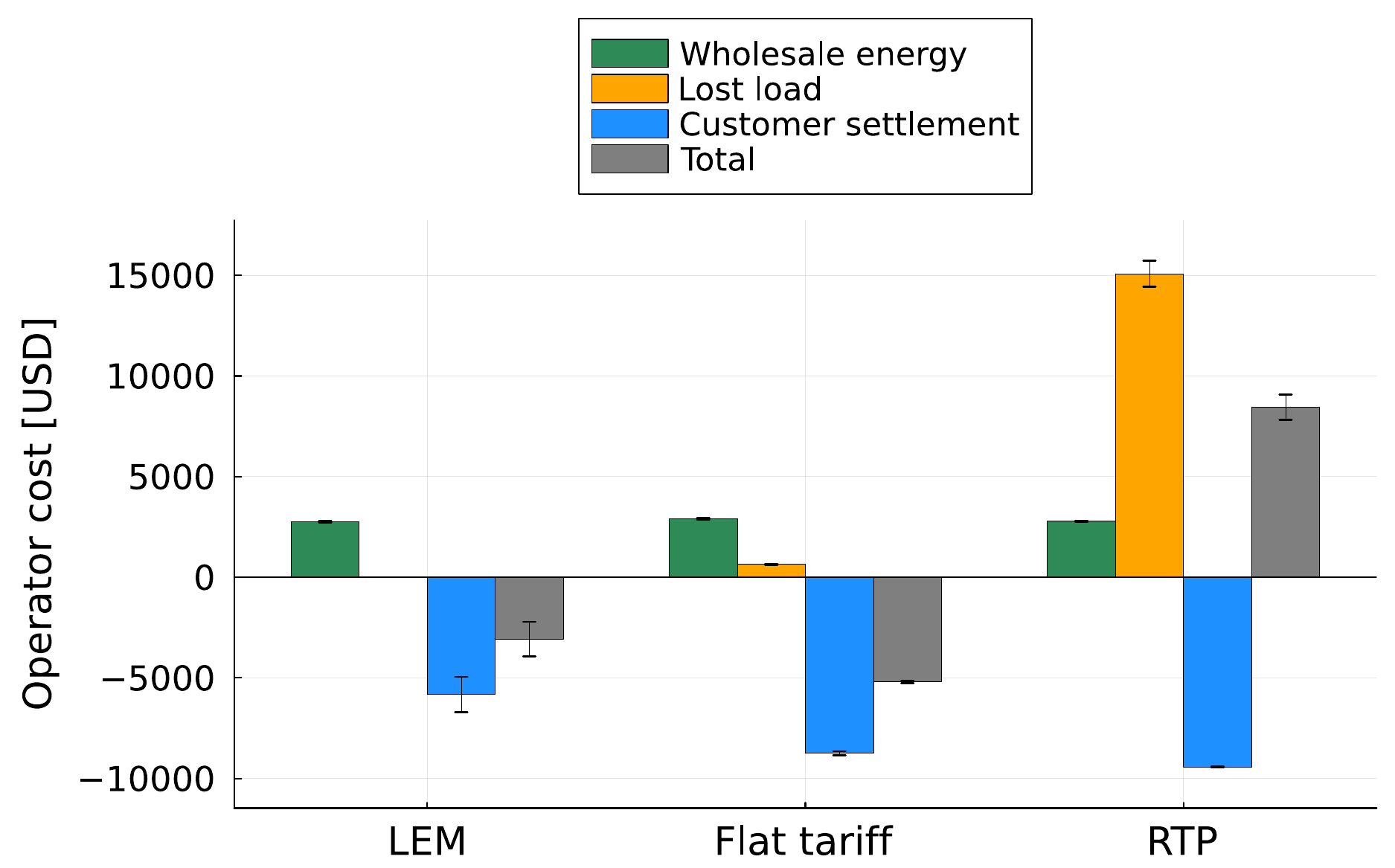}
    \caption{Daily operator cost components. LEM has 41\% higher operator cost compared to FT, and 136\% lower than RTP.}
    \label{fig:operatorcost_50_50}
  \end{subfigure}
  \caption{Cost breakdown by market design (50\% PV, 50\% BS).}
  \label{fig:costs_50_50}
\end{figure*}

Note that system cost is the sum of customer and operator costs -- this disaggregation is useful to understand how the market designs influence different stakeholders. The mean and standard deviation of costs are calculated for a system at 50\% PV penetration and 50\% battery penetration, with the resulting monthly bills shown in \cref{tab:monthly_costs_breakdown}. The system cost components for 10 OU simulations over one day are shown in \cref{fig:systemcost_50_50}, where the LEM has the lowest mean system cost. The LEM has a 22\% lower system cost than the FT and an 85\% lower system cost than RTP. Also, the LEM completely avoids lost load in all runs. By contrast, FT still results in monthly lost-load costs. The imported wholesale energy costs (from the transmission grid) are roughly identical under all three markets, indicating that the differences arise mainly due to changes in lost load and internal transfers within the distribution grid. The 95\% confidence interval is shown for all results and the standard deviations are reported in \cref{tab:monthly_costs_breakdown}. Note that all the results are shown in terms of costs -- negative cost values here indicate a net revenue.

LEM has the lowest overall customer costs among the three markets. Daily customer cost components with 95\% confidence intervals are shown in \cref{fig:customercost_50_50}. The LEM has 33\% lower customer cost compared to FT, and 38\% lower than RTP, respectively. The LEM also shows the largest costs and credits associated with DER operation, i.e. costs for BS charging and credit or revenue for BS discharging and PV generation. This makes sense since such aggressive market-based coordination is what incentivizes customer behavior in the LEM. The LEM shows substantially larger variation across OU runs than RTP and FT. The mean monthly operator costs in both the LEM and FT markets are negative, indicating net revenue, as opposed to the RTP case where operators face net costs. This suggests that RTP schemes may not be commercially favorable for grid operators. FT returns slightly more revenue to operators through customer settlements. LEM has 41\% lower operator revenue compared to FT, but 136\% lower operator cost compared to RTP. \cref{fig:operatorcost_50_50} shows the daily operator cost components together with 95\% confidence intervals. The standard deviations for total operator cost are listed in \cref{tab:monthly_costs_breakdown}. Among the three designs, the LEM shows the largest variation across OU realizations, which is caused by the customer settlement component. We also find that the RTP market is inferior to both LEM and FT in terms of average costs as well as volatilities.

\begin{table*}[t]
\centering
\caption{Monthly cost breakdown by market design. Mean values are shown in bold; standard deviations are shown in parentheses.}
\label{tab:monthly_costs_breakdown}
\small
\begin{tabular}{lcccc c}
\toprule
\textbf{Market} &
\textbf{Operator Costs} &
\textbf{Customer Costs} &
\textbf{Lost Load Costs} &
\textbf{Wholesale Energy Costs} &
\textbf{System Cost} \\
&
\textbf{(\$/month)} &
\textbf{(\$/month)} &
\textbf{(\$/month)} &
\textbf{(\$/month)} &
\textbf{(\$/month)} \\
\midrule
LEM &
\textbf{-92,225} (7,623) &
\textbf{174,902} (7,747) &
\textbf{0} (0) &
\textbf{82,676} (301) &
\textbf{82,676} (301)\\
FT (NEM) &
\textbf{-155,866} (551) &
\textbf{262,381} (890) &
\textbf{19,260} (194) &
\textbf{87,255} (271) &
\textbf{106,515} (333) \\
RTP (NEB) &
\textbf{253,295} (5,595) &
\textbf{282,690} (286) &
\textbf{452,453} (5,634) &
\textbf{83,531} (246) &
\textbf{535,984} (5,639) \\
\bottomrule
\end{tabular}
\end{table*}

\begin{figure}[htbp]
  \centering
  \includegraphics[width=\linewidth]{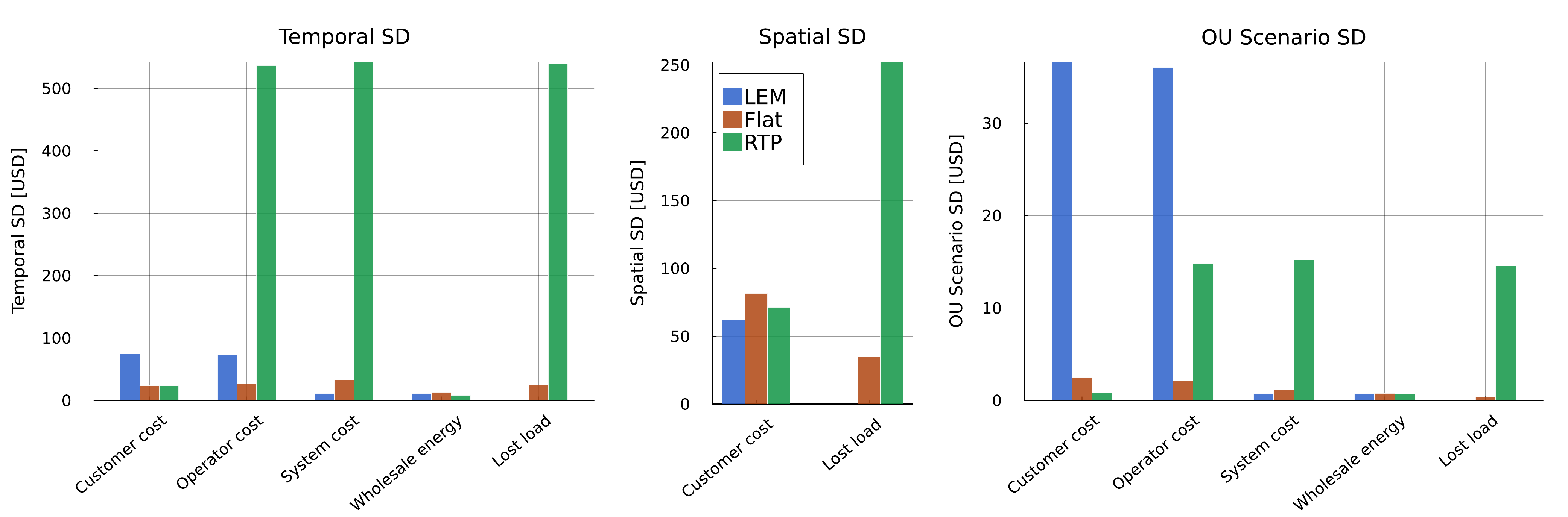}
  \caption{Comparison of volatility or standard deviations (SD) among the three markets. The temporal and spatial SD capture the variability across timesteps and nodes, respectively. Both the temporal and spatial SDs are averaged over the OU runs. The rightmost plot quantifies the variability across different OU process scenarios, averaged over time. Penetration level of 50\% PV and 50\% Battery storage (BS).}
  \label{fig:cost_volatility_compared}
\end{figure}

Looking at \cref{fig:customercost_50_50}, we find that the LEM significantly benefits customers by reducing their expenditure on energy purchased from the grid, while also allowing them to earn more from exporting PV generation and battery discharge. However, they do spend slightly more on battery charging. From the operator perspective in \cref{fig:operatorcost_50_50}, we see that RTP is much worse than both LEM and FT due to massive lost load costs. The FT market is slightly superior to the LEM since it allows the operator to earn more from customer settlements. Although the LEM does lead to some loss in revenue for operators relative to FT, this is outweighed by its reductions in customer costs, resulting in lower system costs overall. \cref{fig:cost_volatility_compared} digs deeper into the cost volatility comparison across the three markets. As expected intuitively, we find that RTP significantly increases temporal volatility as seen in the leftmost plot. Interestingly, we find that this high volatility is primarily driven by the volatility in lost load costs arising from real-time price-based dispatch, rather than the time-varying retail prices themselves. This is also what causes the RTP to have higher node-to-node volatility in lost load costs in the center plot. The spatial volatility in customer costs are similar across the markets, with the FT being most volatile, indicating that the variation is driven by differences in nodal power dispatch. On the other hand, in the rightmost plot, we see that the LEM has much higher uncertainty in operator and customer costs across different OU scenarios or realizations, indicating that it magnifies the effects of forecast errors and uncertainties. This is mainly because PV forecasts strongly influence both PV and battery dispatch, and thus determine internal transfers among the customers and operator. However, system costs are less affected by these internal transfers and here RTP shows the most volatility due to lost load.

\begin{figure}[htbp]
  \centering
  \includegraphics[width=0.7\linewidth]{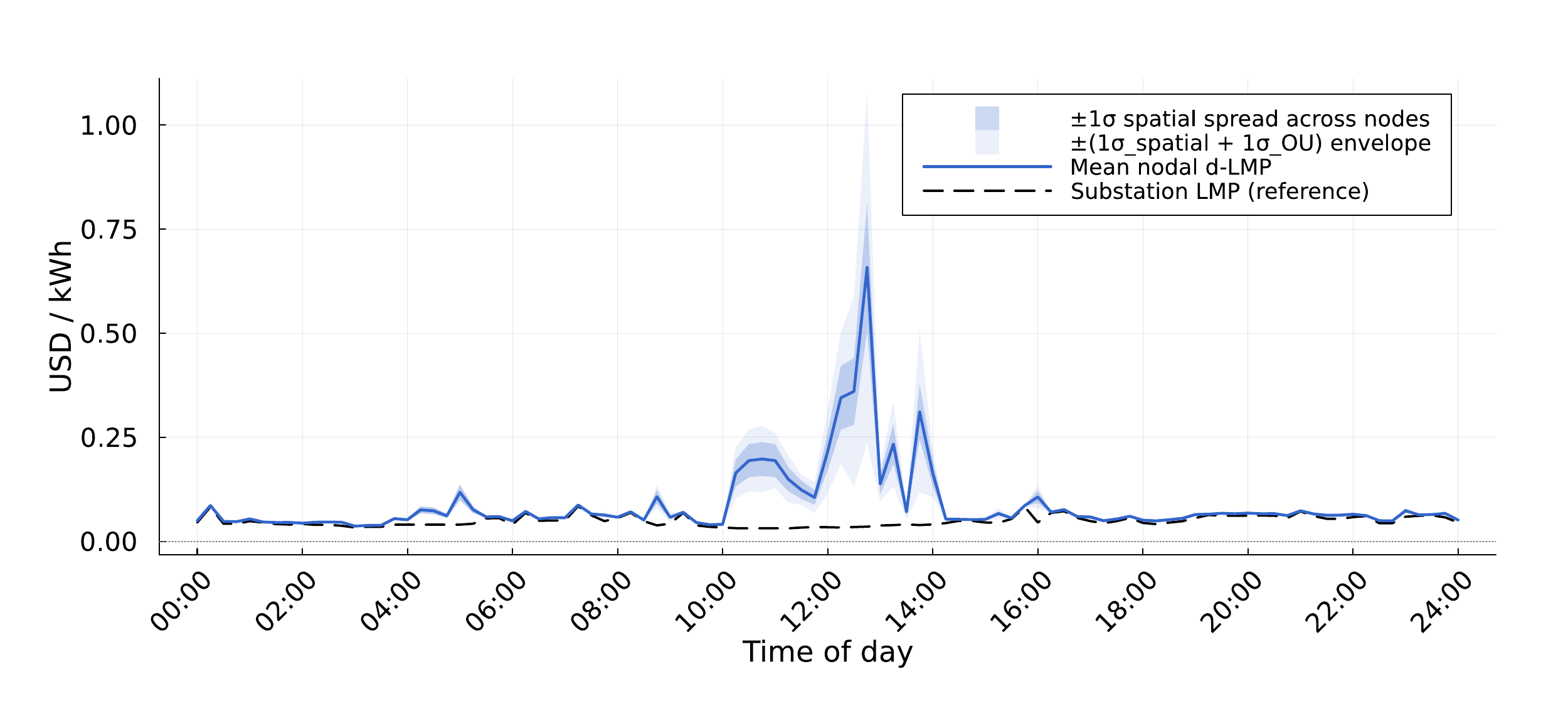}
  \caption{Spatial and OU scenario volatility in d-LMP results from the LEM. (50\% PV, 50\% BS)}
  \label{fig:dlmp_volatility}
\end{figure}

\cref{fig:dlmp_volatility} depicts the volatility in the d-LMPs arising from the LEM over time, compared against the substation node LMP profile as a reference. We find that the d-LMP volatility is generally low and manageable for most periods of the day. However, the volatility significantly increases during the mid-day period (between 10AM-2PM) when the d-LMPs themselves spike as well. This is driven by both the spatial variation in prices across nodes (dark blue shaded region) along with the effects of forecast uncertainty (light blue shaded region). The magnitude of d-LMPs also increases during peak PV output hours around noon, while the LMP remains low. This is likely attributed to high PV output inducing power flow congestion within the distribution grid. This higher price volatility and magnitude needs to be carefully managed by grid operators or regulators since this period also corresponds to peak solar PV output.

\begin{figure}[htbp]
  \centering
  \includegraphics[width=0.6\linewidth]{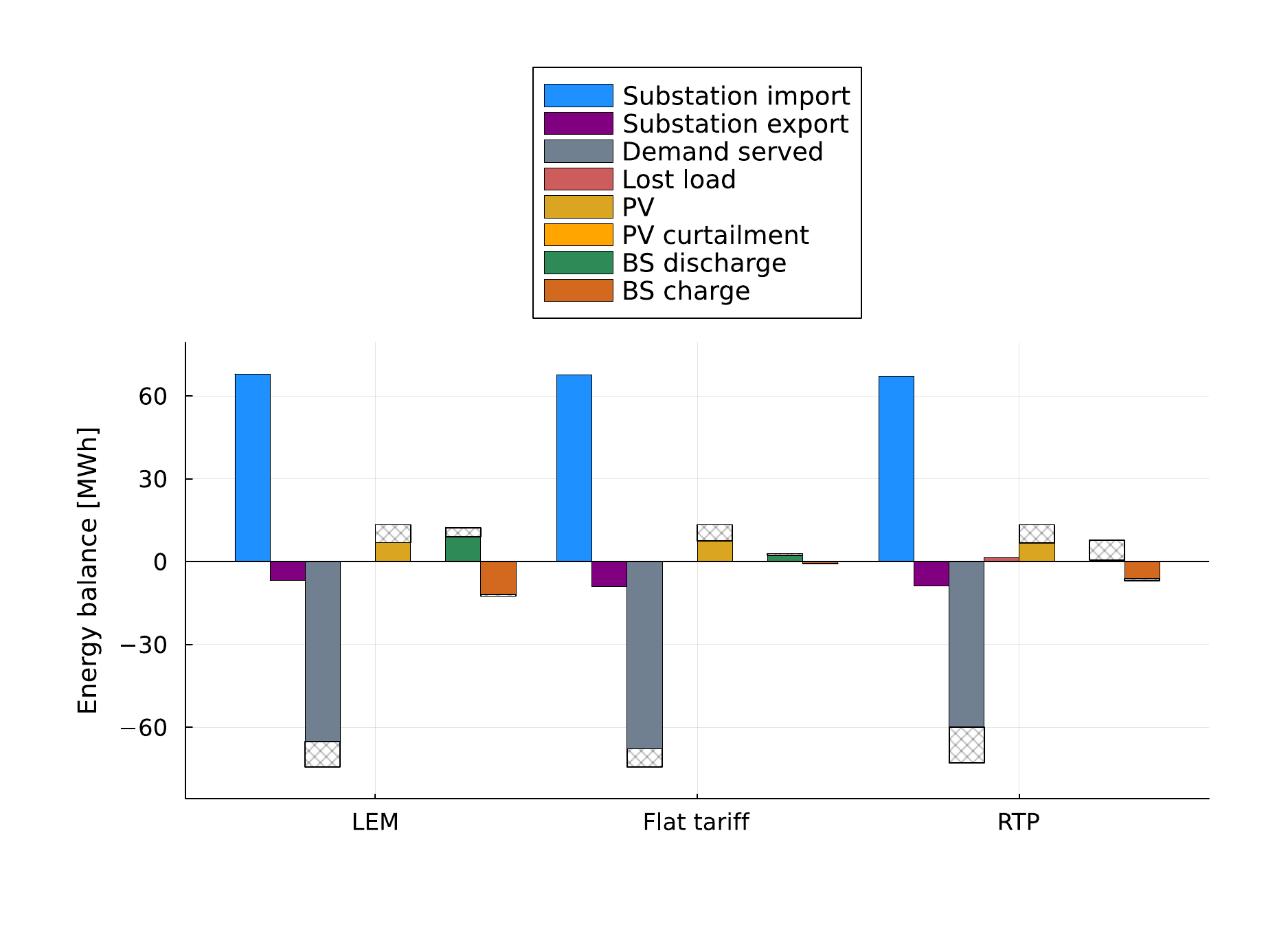}
  \caption{Energy balance. The shaded area shows behind-the-meter internal transfers. (50\% PV, 50\% BS)}
  \label{fig:energy_balance_50_50}
\end{figure}

\cref{fig:energy_balance_50_50} shows the energy balance comparison of all markets. The gray shaded area in each bar shows the behind-the-meter contributions, i.e., the energy flows exchanged among the DERs within the same node (load, PV, batteries). These behind-the-meter flows are calculated using a sequential netting rule in which solar PV serves local demand first and then battery charging, battery discharge serves the remaining local demand, the grid covers any residual deficit, and any remaining surplus is treated as export fro the node back to the grid. All markets have the same electricity demand but the levels of served demand differ due to the non-zero lost load under FT and RTP. It also shows that the LEM has the largest total DER energy flows (shaded and unshaded bar area combined), largely due to higher active participation of batteries in both charging and discharging modes enabled by network-aware coordination. The RTP case shows the highest behind-the-meter contribution (shaded area only), as egoistic optimization combined with the discounted export credit ($\beta = 0.25$) incentivizes DERs to keep energy behind the meter rather than export to the grid.

\subsubsection{Influence of DER penetration\label{sec:der_pen}}
\begin{figure}[htbp]
  \centering
  \includegraphics[width=0.7\linewidth]{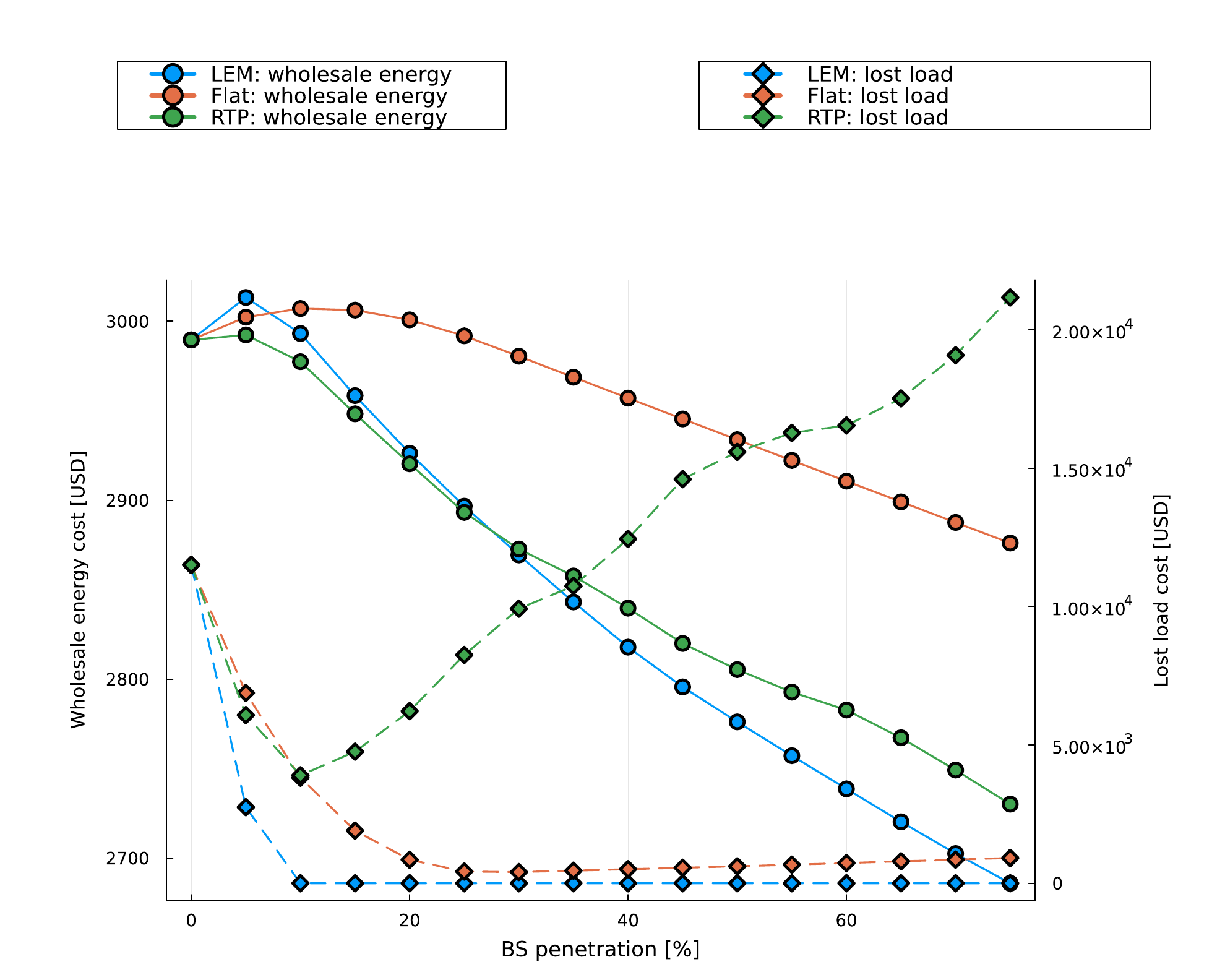}
  \caption{Sensitivity of daily wholesale energy and lost-load cost to battery penetration (PV: 50\%).}
  \label{System_costs_bs_variable}
\end{figure}

Here, we study the influence of DER penetration levels, by focusing on results from just a single OU scenario. This provides a deep dive into how the role played by BS evolves as we vary the penetrations of solar and storage. \cref{System_costs_bs_variable} shows how lost-load and wholesale energy import costs (which together give us system costs), vary with different battery penetrations, at a fixed PV penetration of 50\%. Without any battery penetration, the LEM has the same lost load cost as RTP and FT markets. As battery penetration increases, the LEM eliminates lost load entirely from a penetration of 10\% onwards. In contrast, lost load is not fully eliminated under either FT or RTP over the tested range. As BS penetration rises, the lost load cost remains relatively flat (at a non-zero value) in the FT case, but consistently increases in the RTP case. This surprising result is likely because neither the FT nor RTP markets intelligently optimize or coordinate the BS operation to help reduce lost load, unlike the receding horizon BS dispatch in the LEM. Thus, uncoordinated and suboptimal BS charging operations can potentially worsen lost load in the RTP case, especially if the BS at each node optimizes its operation in a myopic and egoistic manner. Coordinated schemes like the LEM are essential to incentivize batteries at different nodes to coordinate and achieve global, rather than local, optimality. We find that wholesale energy costs generally decrease with increasing BS levels. The FT market has the highest energy costs across most BS penetrations. The RTP market has lower wholesale costs than the LEM until around 30\% BS penetration, above which the LEM shows the lowest energy costs. This implies that multiperiod coordination allows the LEM to more effectively utilize additional storage to minimize expenditure on energy imports from the wholesale market.

\begin{figure}[htbp]
  \centering
  \includegraphics[width=\linewidth]{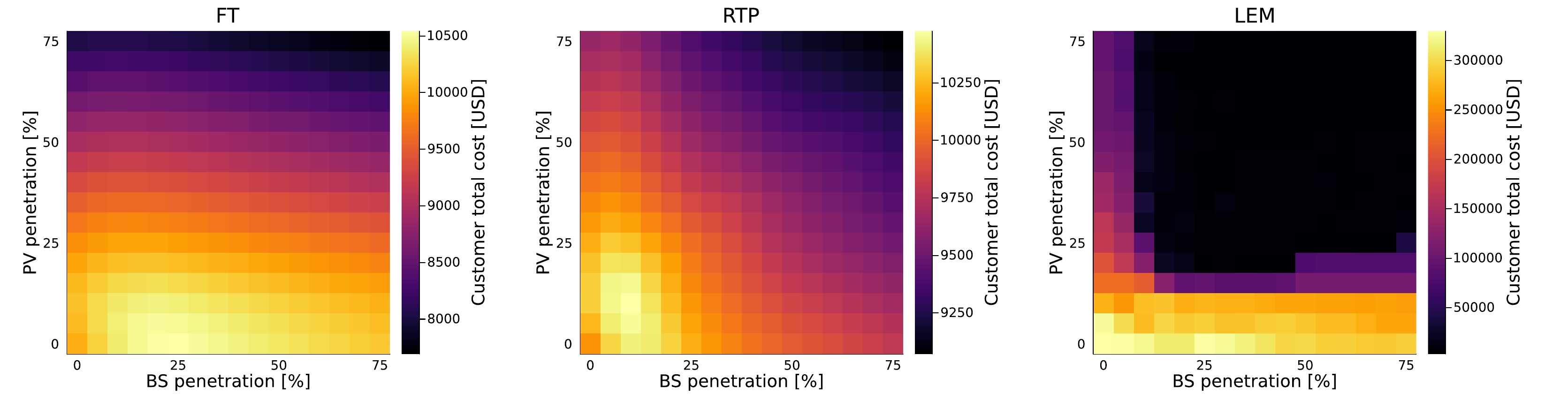}
  \caption{Sensitivity of daily customer cost to PV and battery penetration.}
  \label{fig:heatmap_customer_cost}
\end{figure}

\cref{fig:heatmap_customer_cost} shows how the customer cost varies simultaneously with solar and battery penetrations. 
Under FT, we see that the dominant driver of variation is PV penetration, with higher PV leading to lower customer costs since customers utilize local generation more rather than purchasing grid power. Battery storage has almost no effect with a nearly horizontal gradient, implying that BS adds little to no value under a flat tariff. This makes sense since there are no time-varying signals for the battery to exploit and there is no BS coordination or optimization involved. Under RTP, we find that costs are sensitive to both solar and storage. Both help reduce costs roughly symmetrically and modestly, as we go from the highest costs in the lower left to the lowest costs (high PV + high BS) in the upper right corner. This is because customers locally optimize their own batteries for self-consumption under RTP schemes. In the LEM case, we find that increasing BS and PV penetrations both generally result in lower customer costs. However, the steepest cost reductions occur only when both PV and BS \textit{jointly} increase because the d-LMPs reward flexibility that relieves congestion and batteries dispatch in coordination with grid conditions rather than pure arbitrage. Thus, the LEM is the only market design that truly exploits complementarities between generation and storage assets.

\begin{figure}[htbp]
  \centering
  \includegraphics[width=\linewidth]{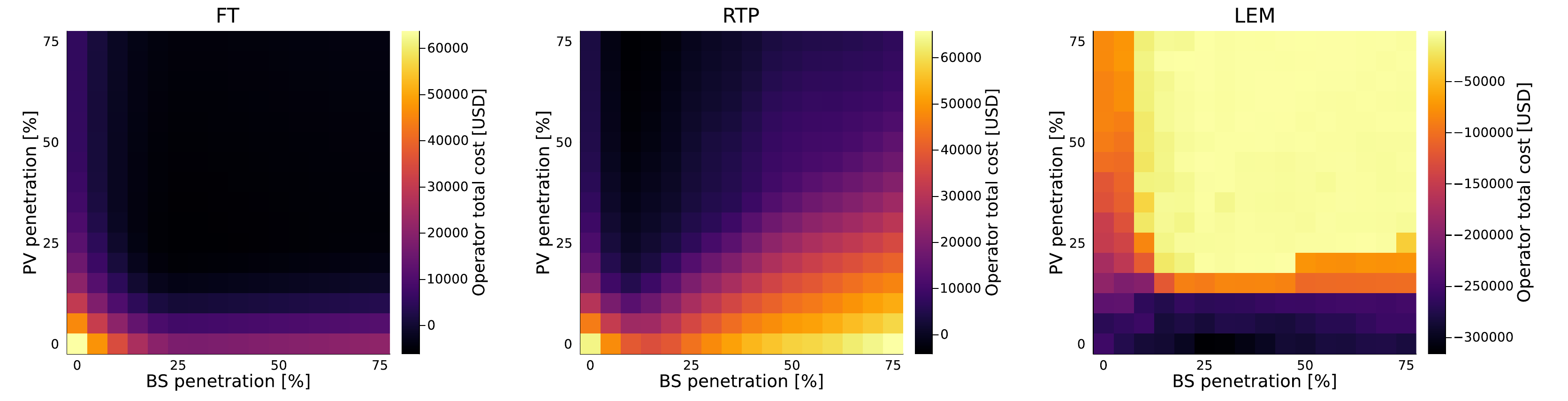}
  \caption{Sensitivity of daily operator cost to PV and battery penetration.}
  \label{fig:heatmap_operator_cost}
\end{figure}

\cref{fig:heatmap_operator_cost} shows how the distribution grid operator cost varies simultaneously with solar and battery penetrations. The first result to note is that the operator costs under the LEM are negative throughout the entire space, indicating net operator revenues. On the other hand, the costs are generally positive for FT and RTP, except at high PV/BS penetrations where operators earn small revenues. Under FT, operator costs generally decline with increasing BS and PV penetration, driven by lower expenditures on wholesale energy costs (from the transmission grid). However, under RTP, we find that operator costs increase with battery penetration (while still decreasing with PV penetration). This is opposite to the effect we saw in customer costs, clearly reflecting the tradeoff between private versus social welfare under RTP. While such egoistic, uncoordinated BS dispatch (arbitrage based on wholesale price signals) relieves customer bills, it also worsens grid conditions for the operator since the network-wide real-time prices don't fully capture distribution-level congestion unlike d-LMPs. Under the LEM, we find that operator revenues are more sensitive to PV penetration, with increasing solar levels leading to lower revenues. However, the operator is still able to maintain modest revenues even at very high PV and BS penetrations approaching 80\%. 

Looking at both sets of customer and operator cost heatmaps, we find that the LEM exhibits a unique, nonlinear, and threshold-driven cost response, unlike with the other two benchmarks where the changes are more gradual and monotonic across the entire range of penetrations. We find that at low PV penetrations (below around 15--20\%), customer costs are significantly higher with the LEM than RTP or FT. This is also accompanied by extremely high operator revenues, indicating a wealth transfer from the customers to the operator via high d-LMPs when the feeder is stressed and flexibility is scarce. However, customer costs drop off a cliff and operator revenues become more modest and reasonable once we go above 15--20\% PV penetration. This is an important policy insight -- LEM performance is highly sensitive to the baseline level of DER deployment. It may underperform FT and RTP in sparse-PV environments, but significantly outperforms them after a critical mass of PV is reached. Thus, LEM implementations should ideally be accompanied by PV deployment support, to avoid poorer outcomes for communities lacking DERs in the short-term. Other protective measures like price caps or competitive safeguards could also be used to reduce the pricing power of operators at low penetrations. Increasing battery penetration alone cannot compensate for insufficient PV. In fact, high BS penetrations at low PV would result in the worst distributional outcomes, with near-peak customer costs and maximum operator profits. This implies that it's crucial to incentivize PV adoption \textit{before} widespread BS deployment. Finally, from the two LEM heatmaps, we notice a Pareto-optimal region in the upper left portion of the penetration space. At high PV penetrations (above 25\%) and low-to-moderate BS penetrations (below 20\%), both the customer and operator heatmaps are simultaneously favorable, with relatively low customer costs and moderate operator profits. Policies targeting this region with a stronger focus on PV in the near-term (and BS later on), will likely result in the most equitable outcomes in terms of balancing both customer and operator welfare. However, note that our analysis only considers monetary aspects (including congestion costs from grid impacts) and does not include other potential benefits of storage, such as backup power and resilience.

\subsubsection{Value of \textit{new} battery capacity in the LEM}

\begin{figure*}[htbp]
  \centering
  \begin{subfigure}[b]{0.49\linewidth}
    \centering
    \includegraphics[width=\linewidth]{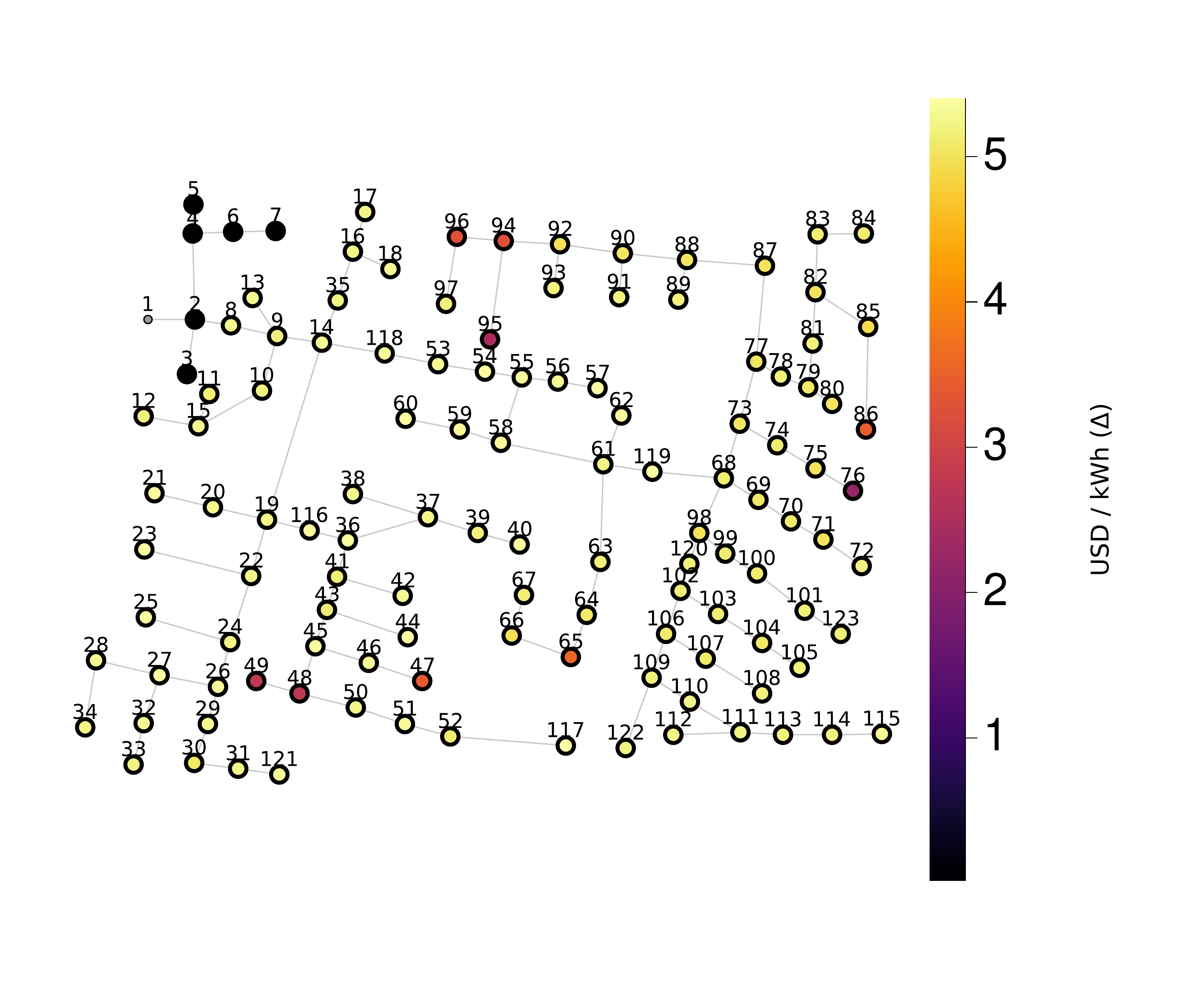}
    \caption{0\% baseline BS.}
    \label{fig:bs_value_20pv_0bs_LEM_private}
  \end{subfigure}
  \hfill
    \begin{subfigure}[b]{0.49\linewidth}
    \centering
    \includegraphics[width=\linewidth]{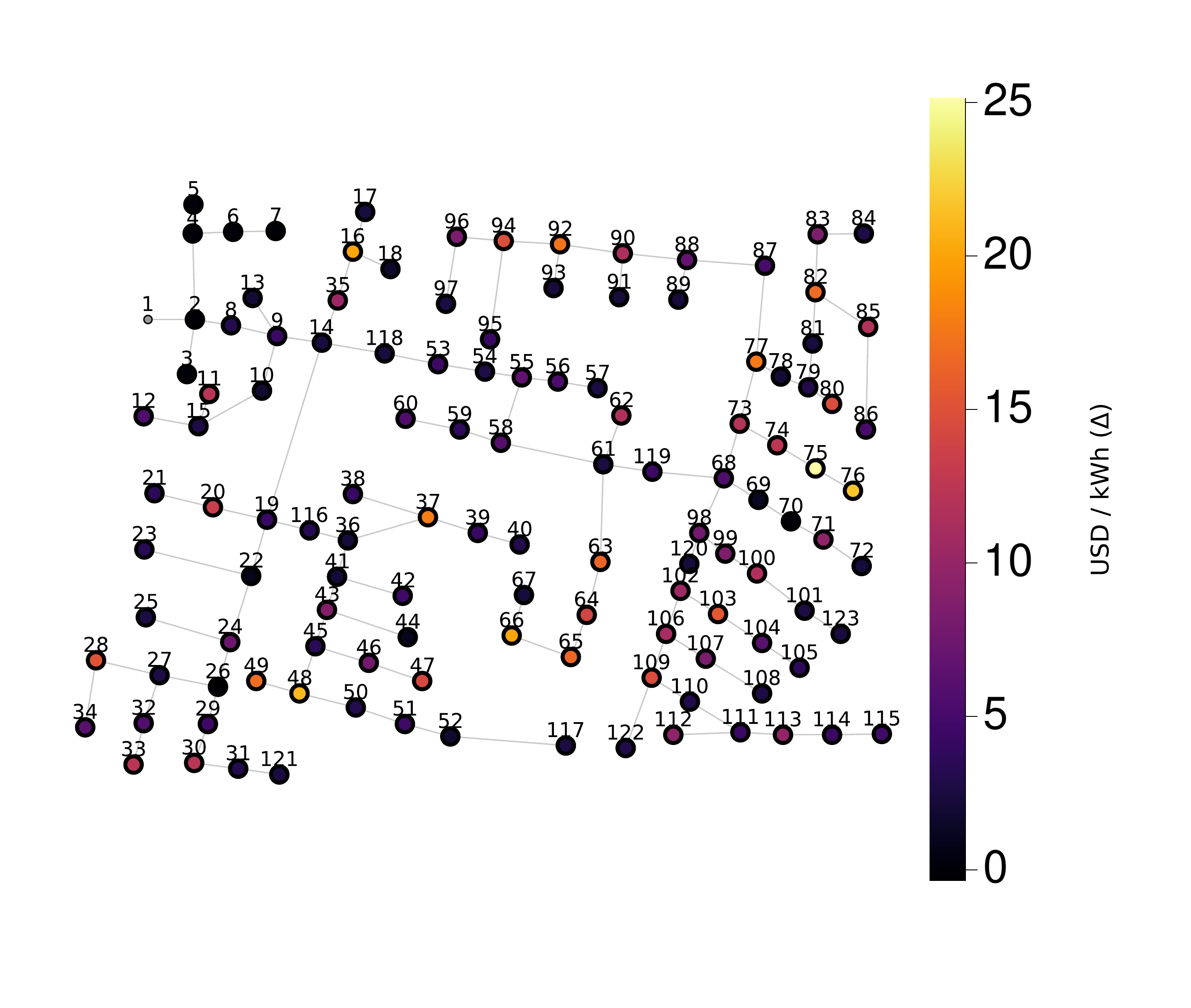}
    \caption{10\% baseline BS.}
    \label{fig:bs_value_20pv_10bs_LEM_private}
  \end{subfigure}
  \hfill
  \begin{subfigure}[b]{0.49\linewidth}
    \centering
    \includegraphics[width=\linewidth]{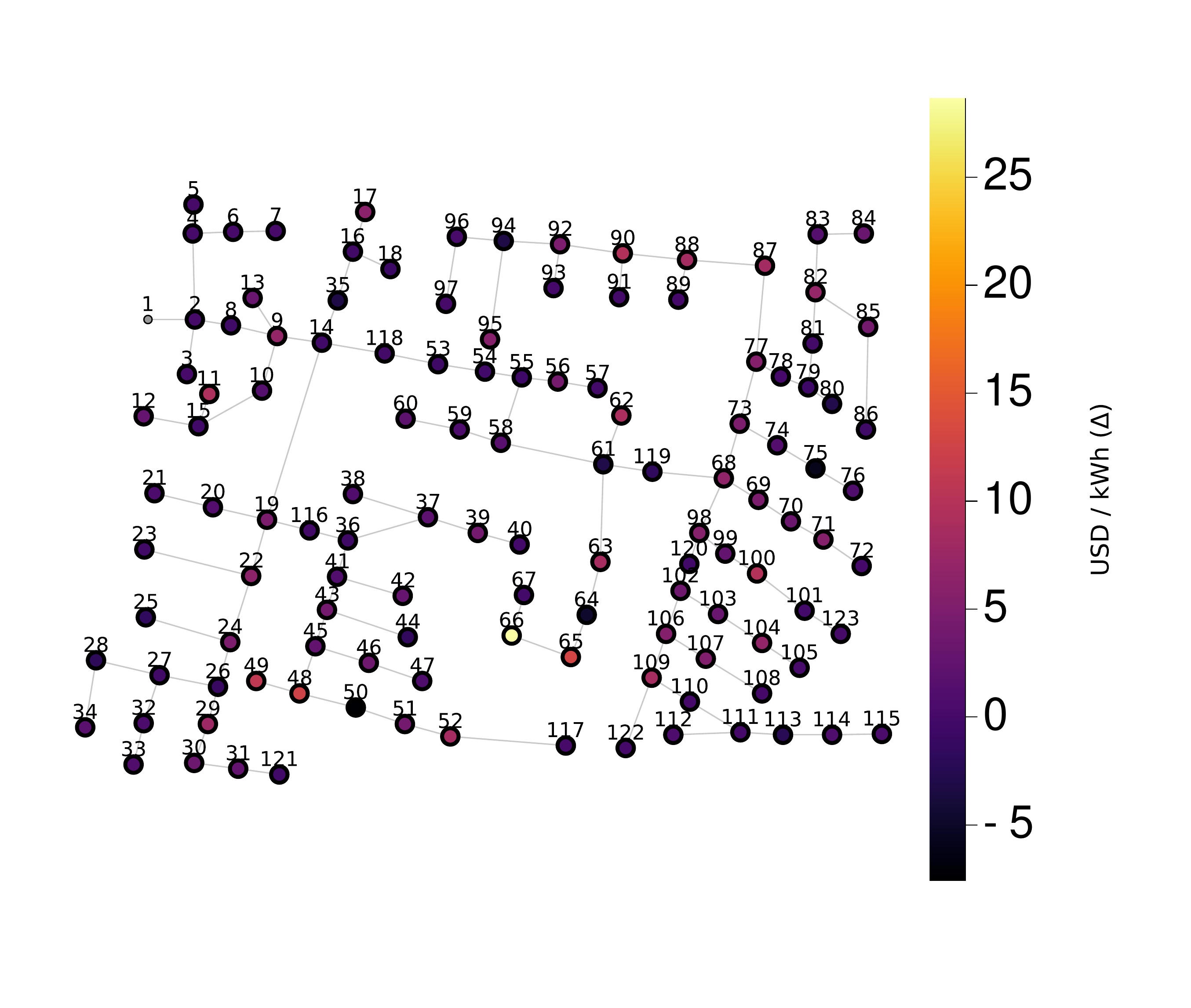}
    \caption{20\% baseline BS.}
    \label{fig:bs_value_20pv_20bs_LEM_private}
  \end{subfigure}
  \hfill
    \begin{subfigure}[b]{0.49\linewidth}
    \centering
    \includegraphics[width=\linewidth]{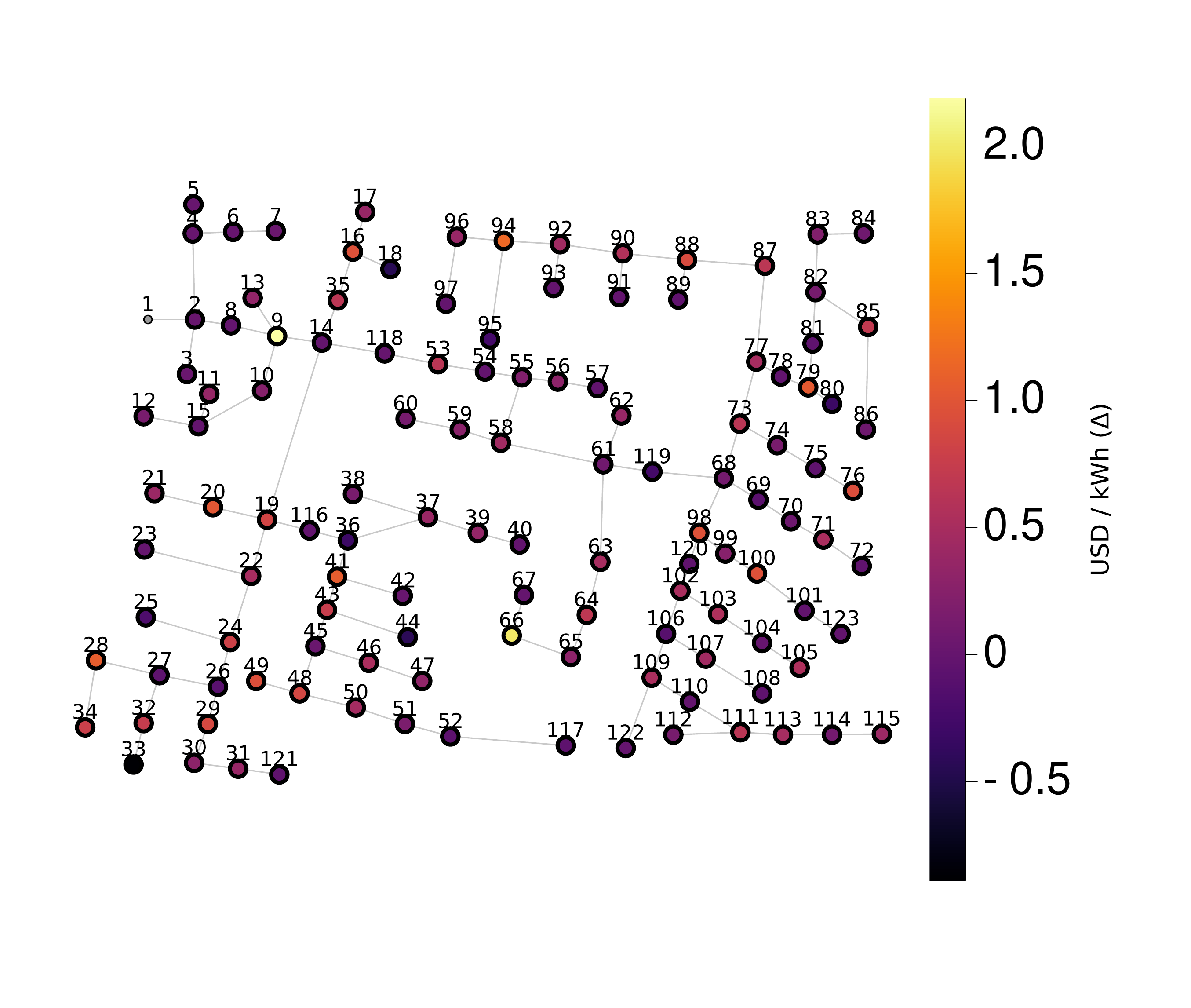}
    \caption{40\% baseline BS.}
    \label{fig:bs_value_20pv_40bs_LEM_private}
  \end{subfigure}
  \caption{Daily marginal private value from increasing battery capacity at node \(i\) (20\% PV).}
  \label{fig:20PV_private_MV}
\end{figure*}

\begin{figure*}[htbp]
  \centering
  \begin{subfigure}[b]{0.49\linewidth}
    \centering
    \includegraphics[width=\linewidth]{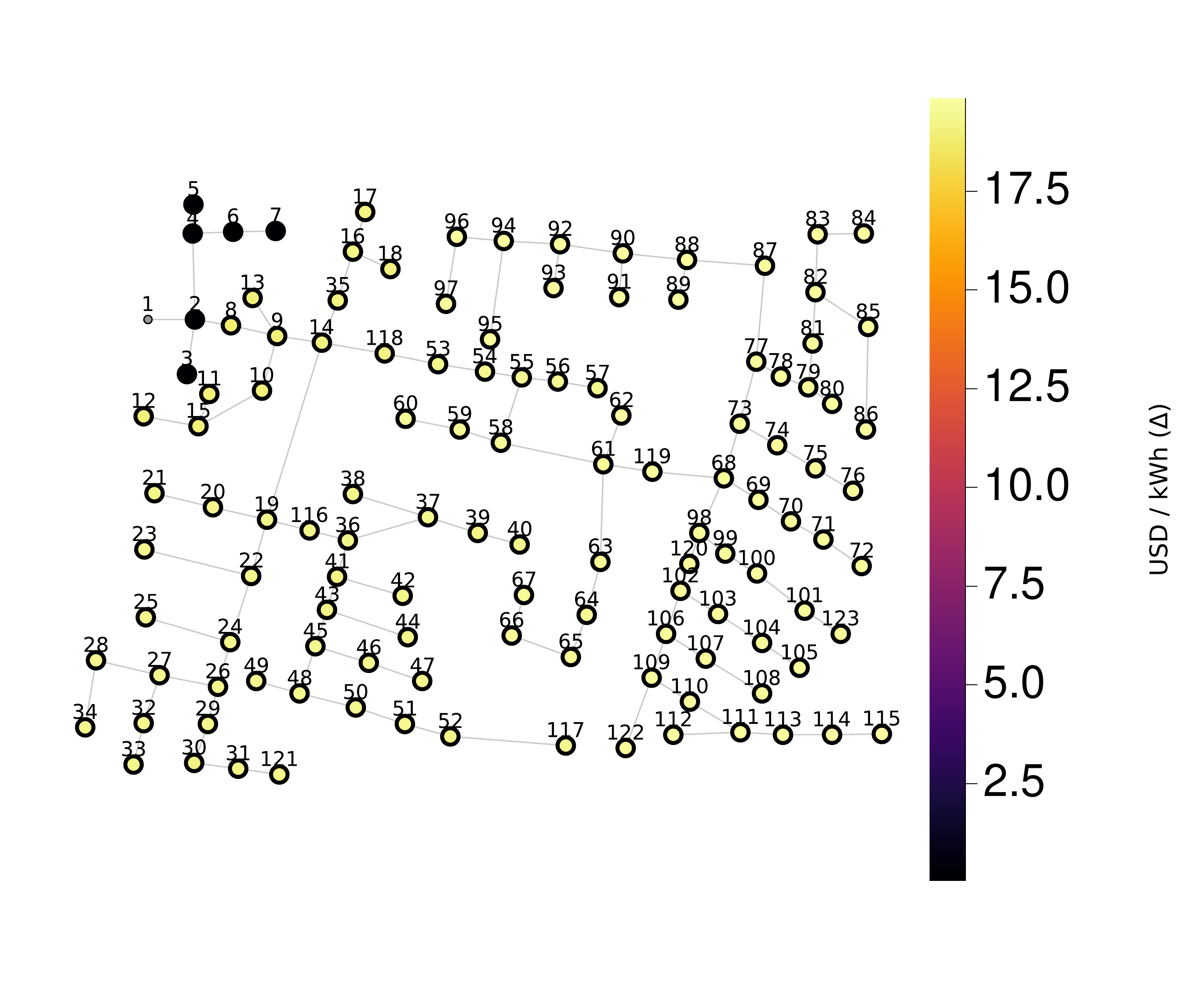}
    \caption{0\% baseline BS.}
    \label{fig:bs_value_20pv_0bs_LEM_system}
  \end{subfigure}
  \hfill
  \begin{subfigure}[b]{0.49\linewidth}
    \centering
    \includegraphics[width=\linewidth]{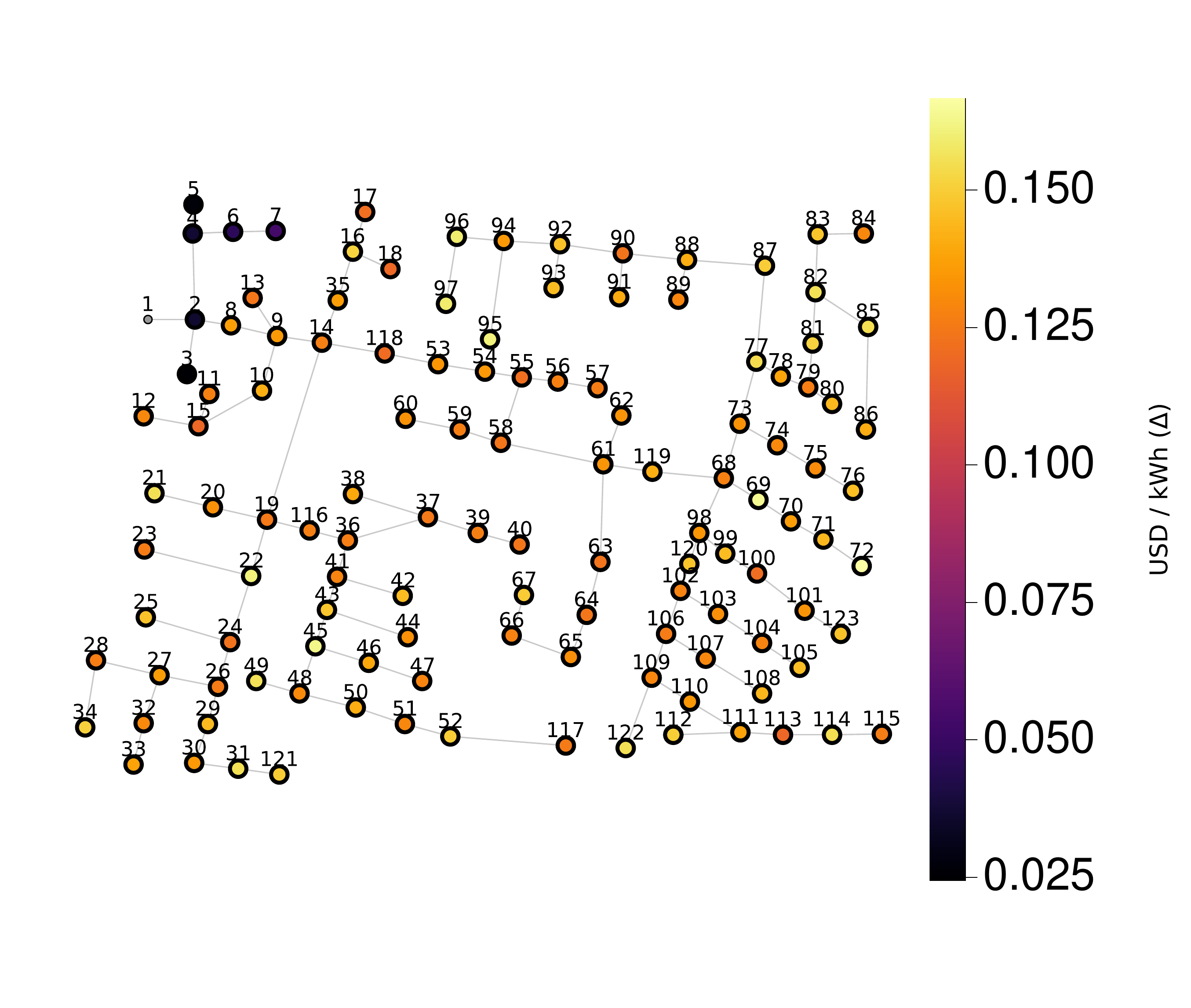}
    \caption{10\% baseline BS.}
    \label{fig:bs_value_20pv_10bs_LEM_system}
  \end{subfigure}
  \hfill
  \begin{subfigure}[b]{0.49\linewidth}
    \centering
    \includegraphics[width=\linewidth]{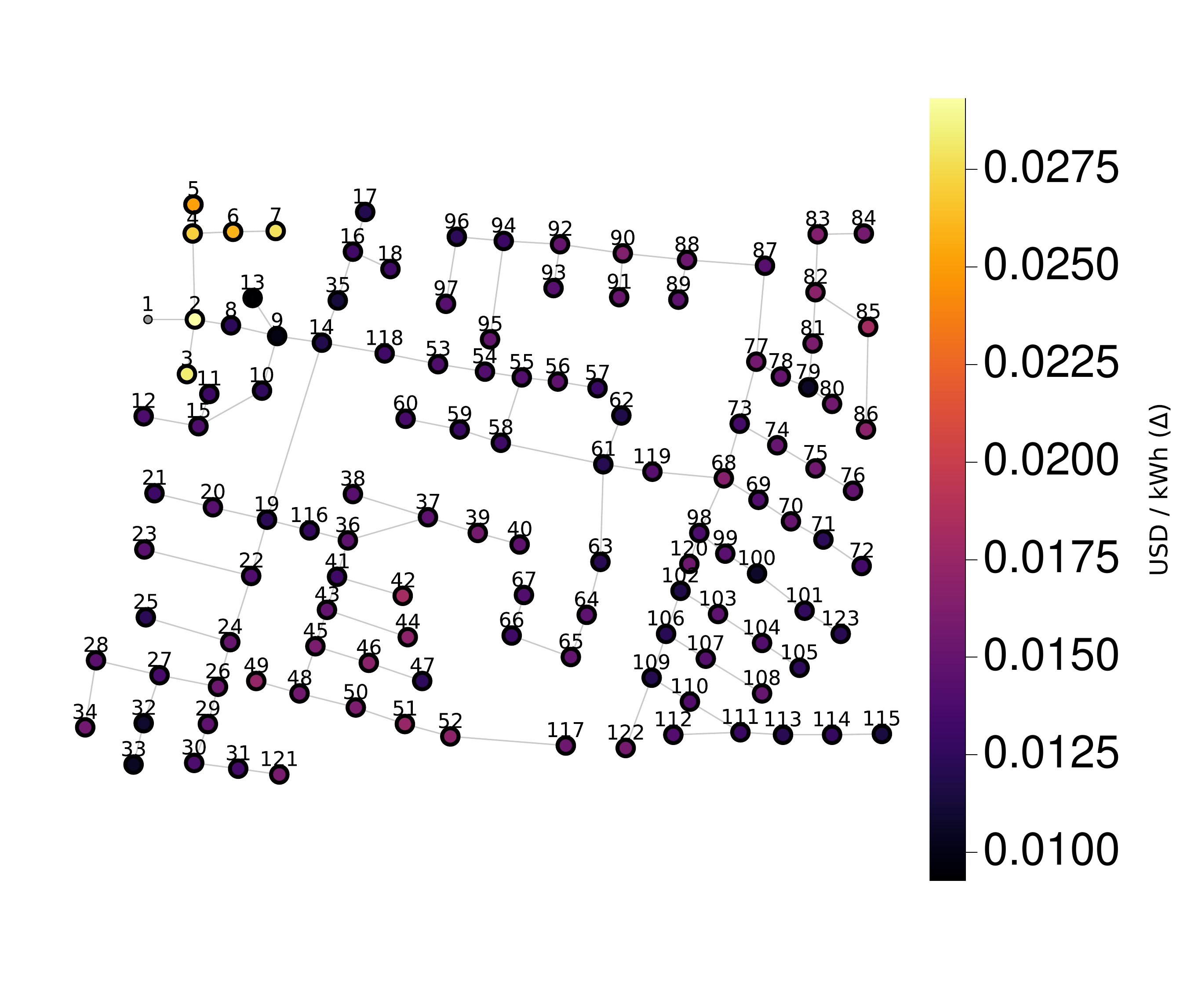}
    \caption{20\% baseline BS.}
    \label{fig:bs_value_20pv_20bs_LEM_system}
  \end{subfigure}
  \hfill
  \begin{subfigure}[b]{0.49\linewidth}
    \centering
    \includegraphics[width=\linewidth]{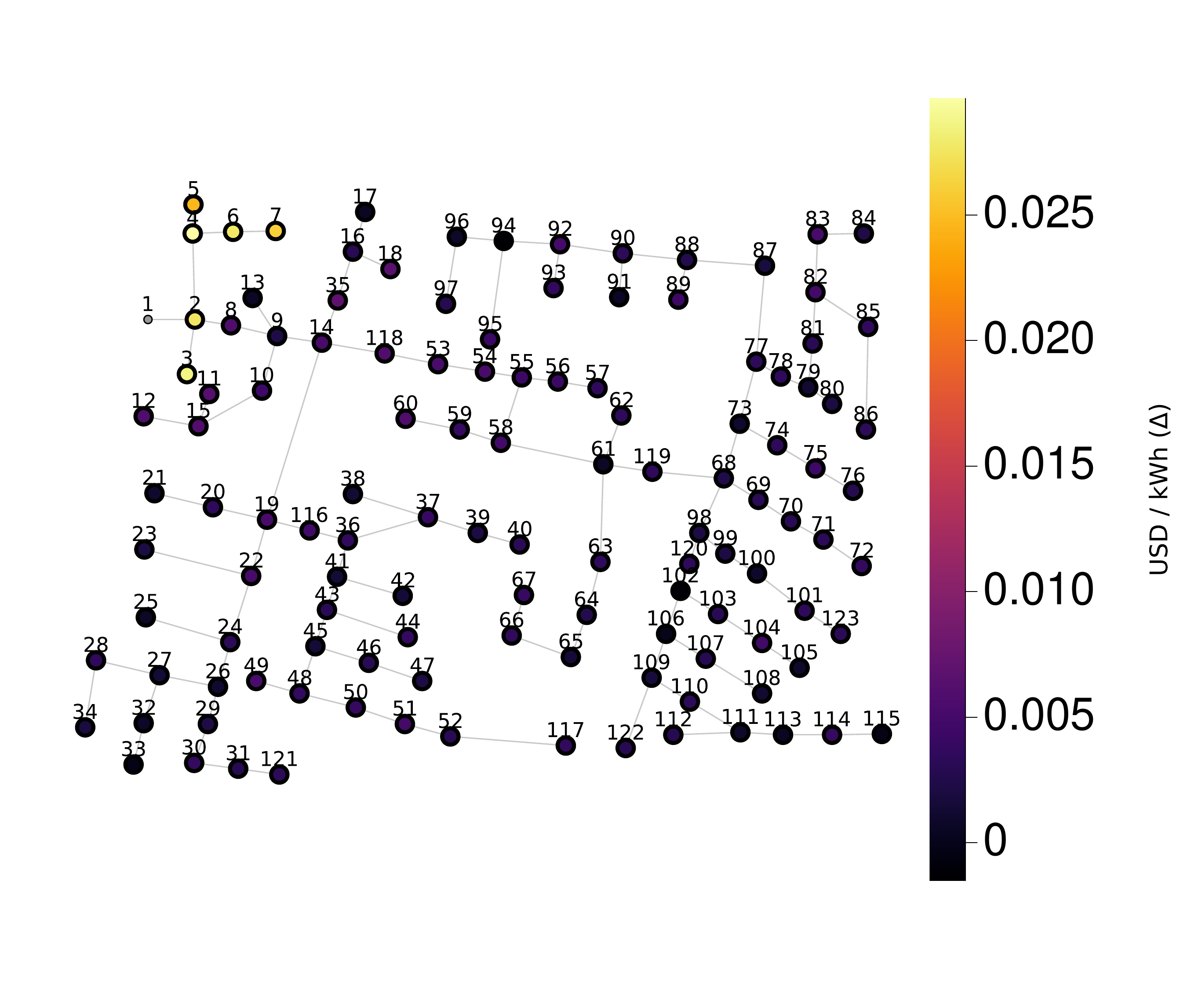}
    \caption{40\% baseline BS.}
    \label{fig:bs_value_20pv_40bs_LEM_system}
  \end{subfigure}
  \caption{Daily marginal system value from increasing battery capacity at node \(i\) (20\% PV).}
  \label{fig:20PV_system_MV}
\end{figure*}

The marginal value of additional battery storage is evaluated using the method described in \cref{BSMV}, for each bus in the test feeder. Note that for the rest of the paper, we focus solely on the LEM. \cref{fig:20PV_private_MV} shows the daily change in private customer costs when battery capacity is increased at a single node, starting from different baseline penetrations, ranging from no initial BS penetration, up to a 40\% baseline BS penetration. \cref{fig:20PV_system_MV} shows the same analysis but for the changes in system costs. In both cases, PV penetration is fixed at 20\%, and the DER siting is the same as in \cref{fig:ieee123_backbone}. The results report the aggregate values over the modeled day, attributed to the perturbed node in the grid. \cref{tab:marginal_value_20PV} shows the mean marginal values and the standard deviation across nodes for all the tested cases.

At zero initial battery penetration, additional battery capacity reduces both private cost and system cost across the feeder. The largest private-cost reduction is observed at node 57, with a marginal value of 5.4 \$/kWh/day, as shown in \cref{fig:20PV_private_MV}, while the largest system-cost reduction is observed at node 62, with a system marginal value of 19.8 \$/kWh/day, as shown in \cref{fig:20PV_system_MV}. Across the downstream part of the feeder, at the nodes 8--123, the system value of added storage is relatively similar. Looking at \cref{fig:20PV_system_MV} and \cref{tab:marginal_value_20PV}, we find that the system MV drops precipitously when we jump from no batteries even to a low baseline penetration of 10\%, and continues to drop further. This is because at low battery penetrations, each additional unit of storage has much more value in reducing scarcity, relieving congestion and depressing d-LMPs. We also observe similarly diminishing BS private MV as we increase baseline BS penetration. However, there's an exception where going from 0 to 10\% slightly increases the average private MV but also significantly increases the volatility and outliers, indicating that at low-to-moderate BS penetrations, most of the private MV is concentrated at a few nodes in the network. Above moderate penetrations (20\% and higher) the private MVs consistently decline with increasing baseline BS. For example, at 20\% initial battery penetration, the value of additional storage becomes smaller and more dependent on the specific nodal location in the network. The largest private cost reduction is observed at node 66, with a marginal private value of 28.7~\$/kWh/day, which is more than twice as high as the second-highest value, as shown in \cref{fig:20PV_private_MV}. Largest system-cost reduction is observed at node 2, with a marginal system value of 0.029~\$/kWh/day, as shown in \cref{fig:20PV_system_MV}. In fact, at higher penetrations, we start to observe that both private and system MV of storage can become negative at several nodes. This also connects with our previous results in \cref{sec:der_pen}, where we saw that increasing BS deployment beyond a certain level can potentially worsen outcomes for both customers and operators.

\begin{figure*}[htbp]
  \centering
  \begin{subfigure}[b]{0.49\linewidth}
    \centering
    \includegraphics[width=\linewidth]{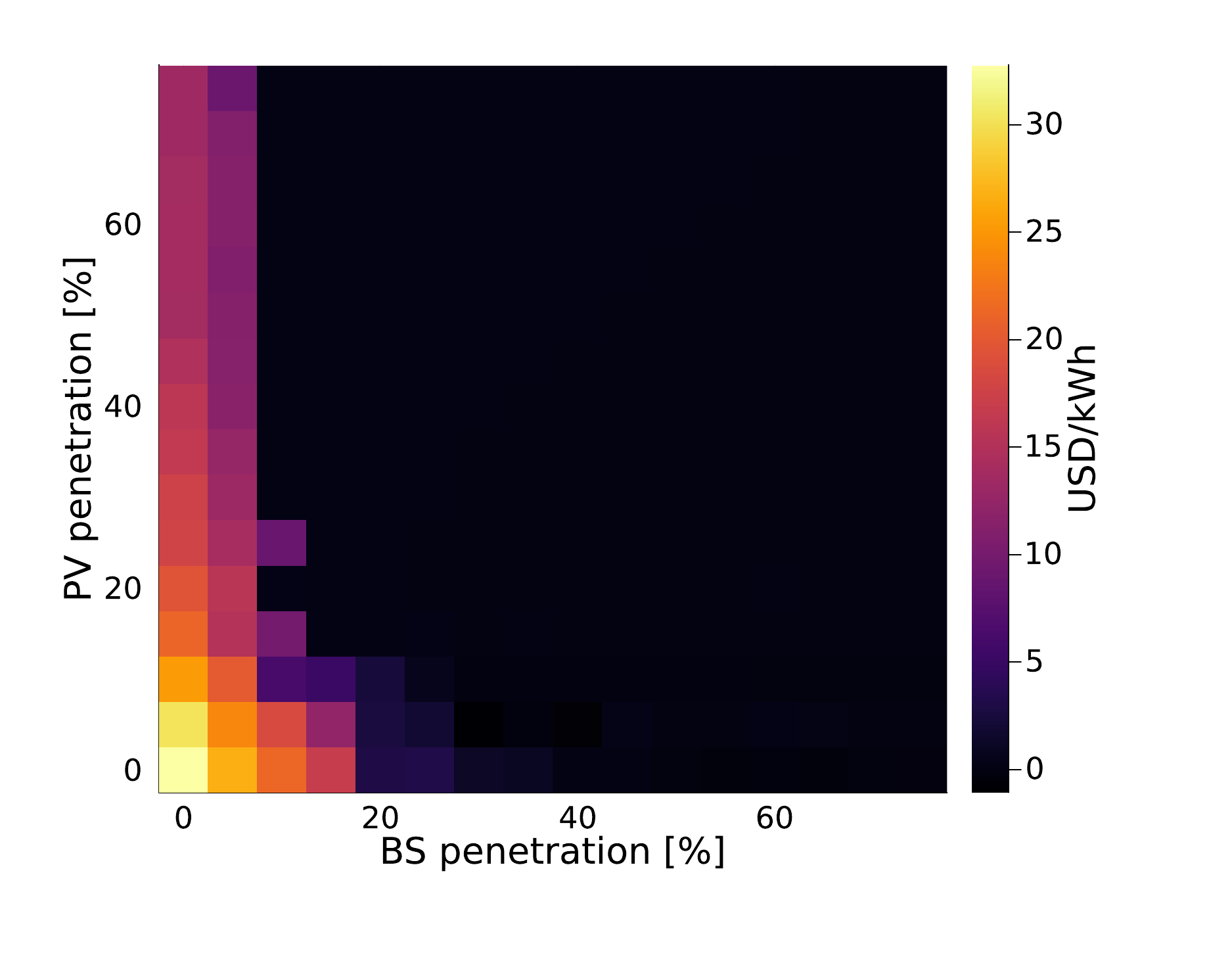}
    \caption{System marginal value.}
    \label{fig:systemMV_node61}
  \end{subfigure}
  \hfill
  \begin{subfigure}[b]{0.49\linewidth}
    \centering
    \includegraphics[width=\linewidth]{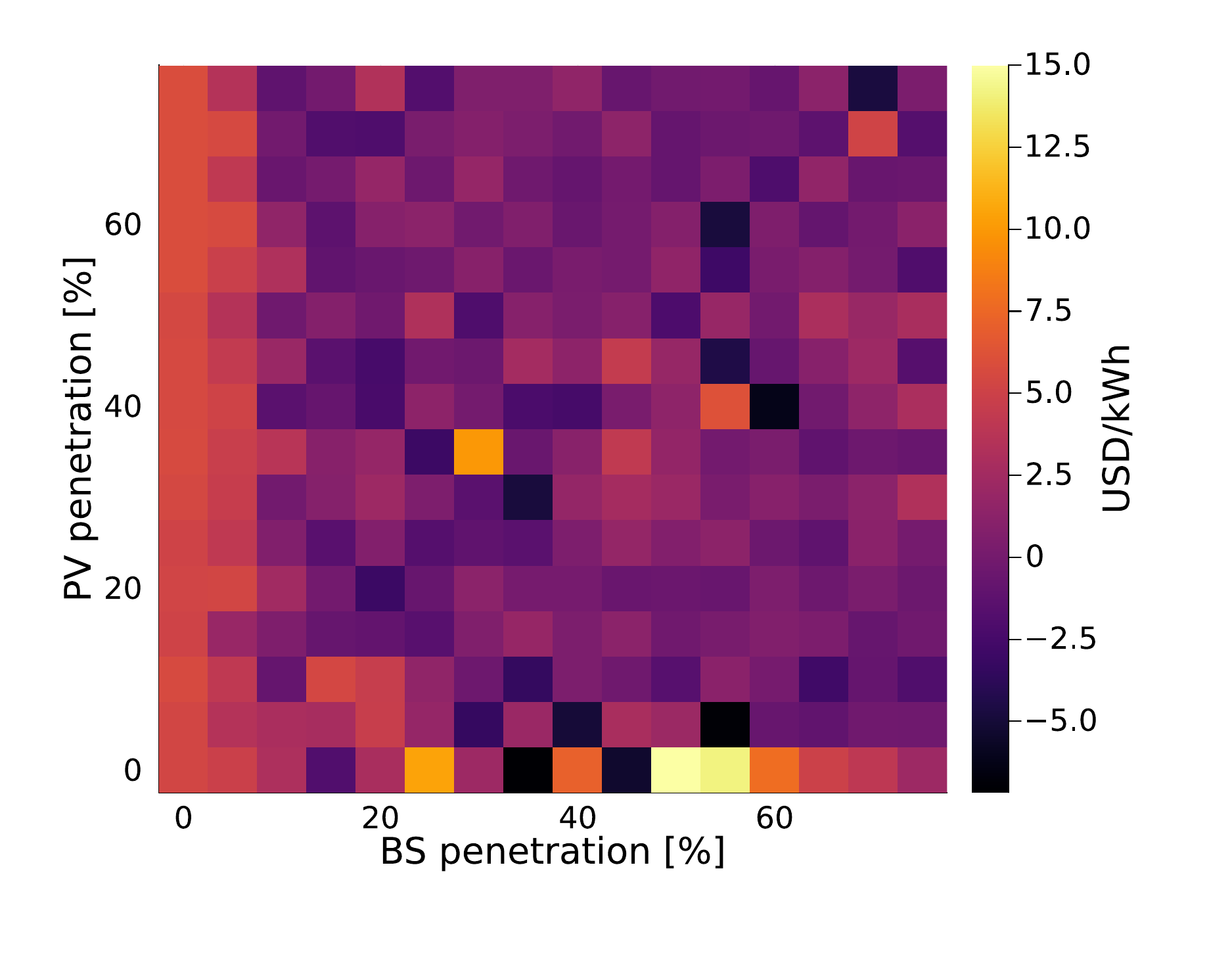}
    \caption{Private marginal value.}
    \label{fig:privateMV_node61}
  \end{subfigure}
  \caption{Daily marginal value from increasing battery capacity at node 61, with different PV and battery penetrations.}
  \label{fig:MV_node61}
\end{figure*}
\begin{figure}[htbp]
    \centering
    \begin{subfigure}[b]{0.49\linewidth}
    \includegraphics[width=\linewidth]{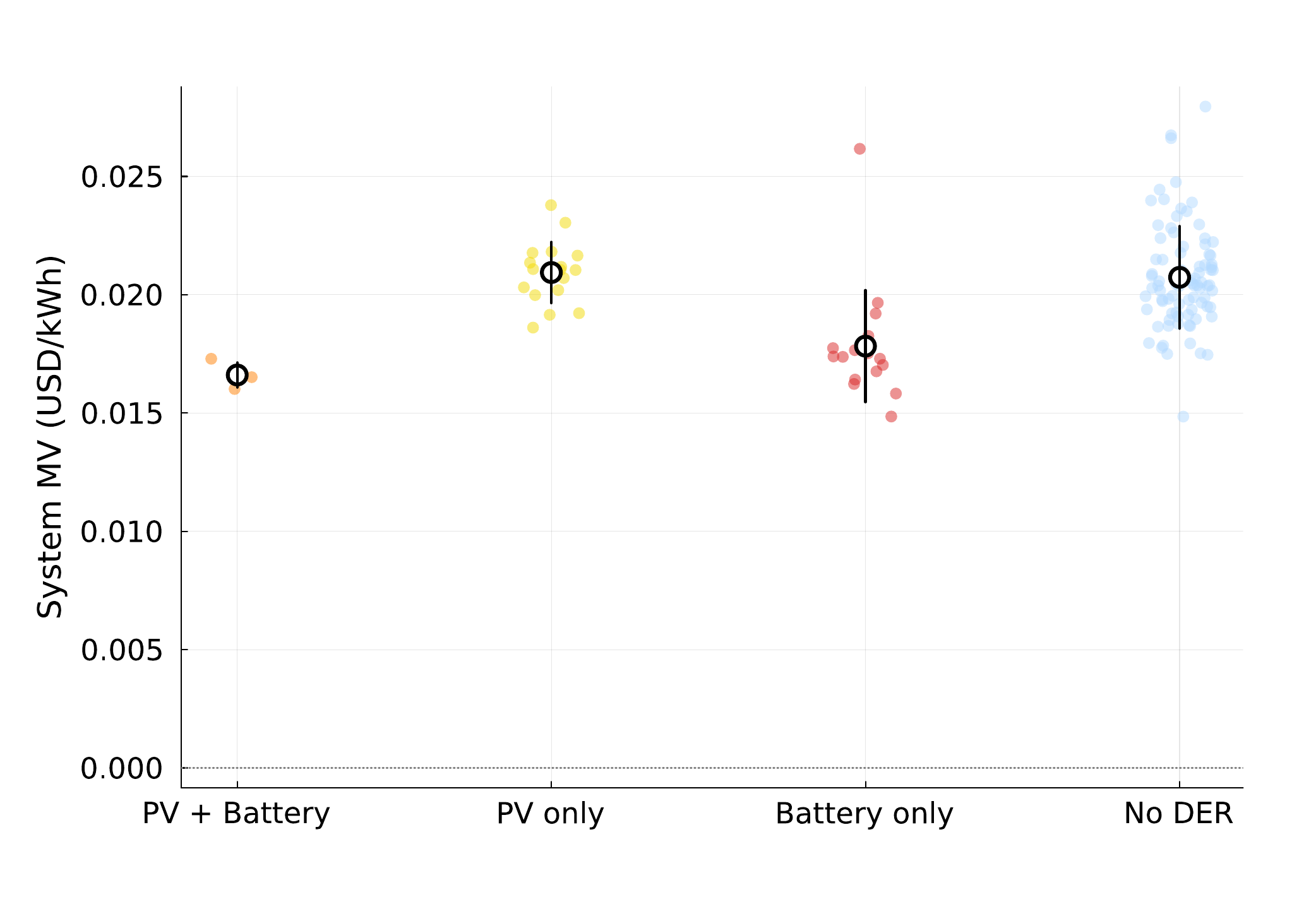}
    \caption{System marginal value by bus type (20\% PV, 20\%BS). \label{fig:systemMV_bustype}}
    \end{subfigure}
    \begin{subfigure}[b]{0.49\linewidth}
    \includegraphics[width=\linewidth]{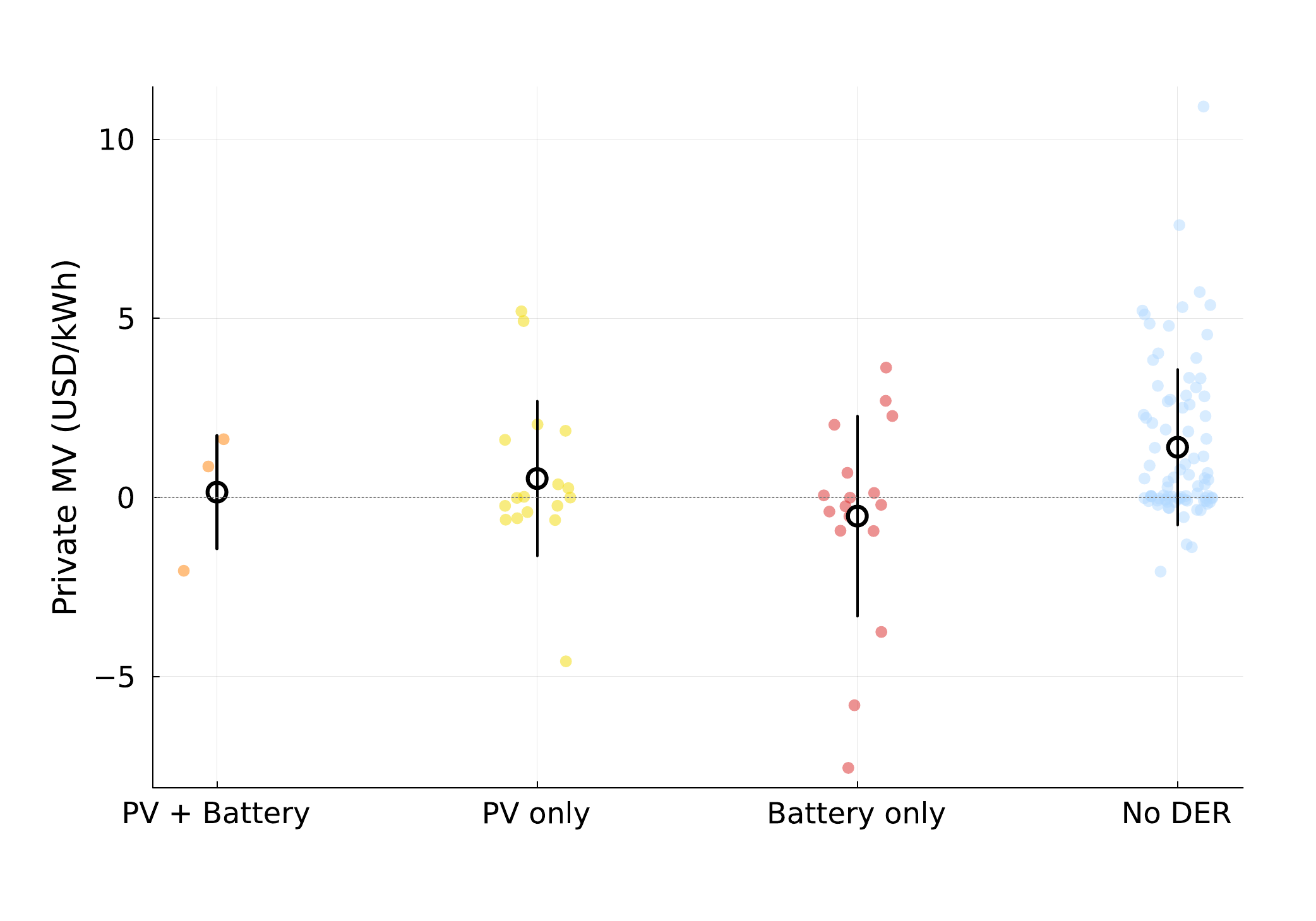}
    \caption{Private marginal value by bus type (20\% PV, 20\%BS). \label{fig:privateMV_bustype}}
    \end{subfigure}
    \caption{Distributions of system and private marginal values of added storage depending on the type of bus, showing the mean value $\pm$ one standard deviation.\label{fig:MV_bustype}}
\end{figure}

Overall, the results show that added storage has higher value when initial battery penetration is low, and that location becomes increasingly important as baseline battery penetration rises. This is clearly shown in \cref{fig:MV_node61} where battery capacity was increased at node 61, a centrally located node in the feeder, with different baseline DER penetrations of both solar and batteries. Following from the trends observed above, we see that both private and system MVs tend to decline with increasing BS penetration. Interestingly, we find that the MV of storage also declines with increasing PV penetration. As the amount of surplus local generation from solar increases, the value of flexibility from additional storage decreases since PV can also play similar (and sometimes redundant) roles in reducing net load and managing grid constraints. While the decline in system MV is quite rapid and monotonic throughout the entire penetration space, the changes in private MV are more gradual and non-uniform, suggesting that these depend more intricately on the specific combinations of PV and BS penetrations being considered. However, we still observe values declining generally as we go from the lower left to the upper right corners of the heatmap. \cref{fig:MV_bustype} differentiates the system and private MV of storage based on the types of DERs present at each node. We find that the system MV of new BS capacity tends to be slightly higher when the BS is either colocated with solar PV, or when sited at nodes which previously had no DERs (BS or PV), while the private MV of new BS is also highest on average at nodes without any DERs.

\begin{table*}[h!]
\centering
\caption{Mean and standard deviation of private and system marginal value (MV) of batteries across nodes under different baseline storage penetrations. Values are reported in USD/kWh/day and all cases use the same PV penetration of 20\%.}
\label{tab:marginal_value_20PV}
\small
\begin{tabular}{llcccc}
\toprule
\textbf{MV type} &
\textbf{Statistic [\$/kWh/day]} &
\textbf{0\% baseline BS} &
\textbf{10\% baseline BS} &
\textbf{20\% baseline BS} &
\textbf{40\% baseline BS} \\
\midrule
Private MV
& Mean              & 4.8  & 6.7    & 2.3    & 0.3    \\
& Standard deviation & 1.2  & 5.8    & 4.4    & 0.4    \\
\midrule
System MV
& Mean              & 18.5 & 0.1321 & 0.0147 & 0.0042 \\
& Standard deviation & 4.2  & 0.0247 & 0.0034 & 0.0056 \\
\bottomrule
\end{tabular}
\end{table*}


\FloatBarrier

\subsection{Value of \textit{existing} battery capacity in the LEM}

Finally, we present another potential approach to value battery storage based on our optimization formulation, mirroring dual-based electricity pricing. We extract the dual variables or shadow prices corresponding to the battery's state of charge update constraint in \cref{eq:soc_update} at each node. The duals of this equality constraint will generally be non-zero and reflect the impacts of maintaining the SOC dynamics on the objective function. We term this the SOC value. Similarly, we also extract the dual solutions corresponding to the battery's power capacity limits (upper bounds) at each node -- this is the power cap dual. We perform appropriate unit conversions to convert the SOC value and power cap duals to be in terms of \$/MWh and \$/MW, respectively. Note that the mean and standard deviation of the power cap duals are very small since these duals corresponding to the upper charging or discharging power limits of the battery are usually zero unless the inequality constraints are binding or active (i.e., hold with equality). We then perform a sensitivity analysis to study how both these duals vary with BS and PV penetrations. Note that this dual-based approach captures something conceptually quite different from the marginal value-based method above. The SOC dual is the shadow price on the battery energy constraint within a fixed fleet -- it answers: ``given the batteries already installed, how much is one additional MWh of stored energy worth to the system?'' On the other hand, the marginal value of BS answers a different question: ``how much does total system (or customer) cost fall if we install more battery capacity to the fleet?''. Thus, conceptually the SOC dual represents the value of energy within the existing storage fleet (i.e., the \textit{intra}-fleet scarcity price) while the MV represents the value of expanding the storage fleet (i.e., \textit{inter}-fleet diminishing returns).

\begin{figure}[htbp]
  \centering
  \includegraphics[width=\linewidth]{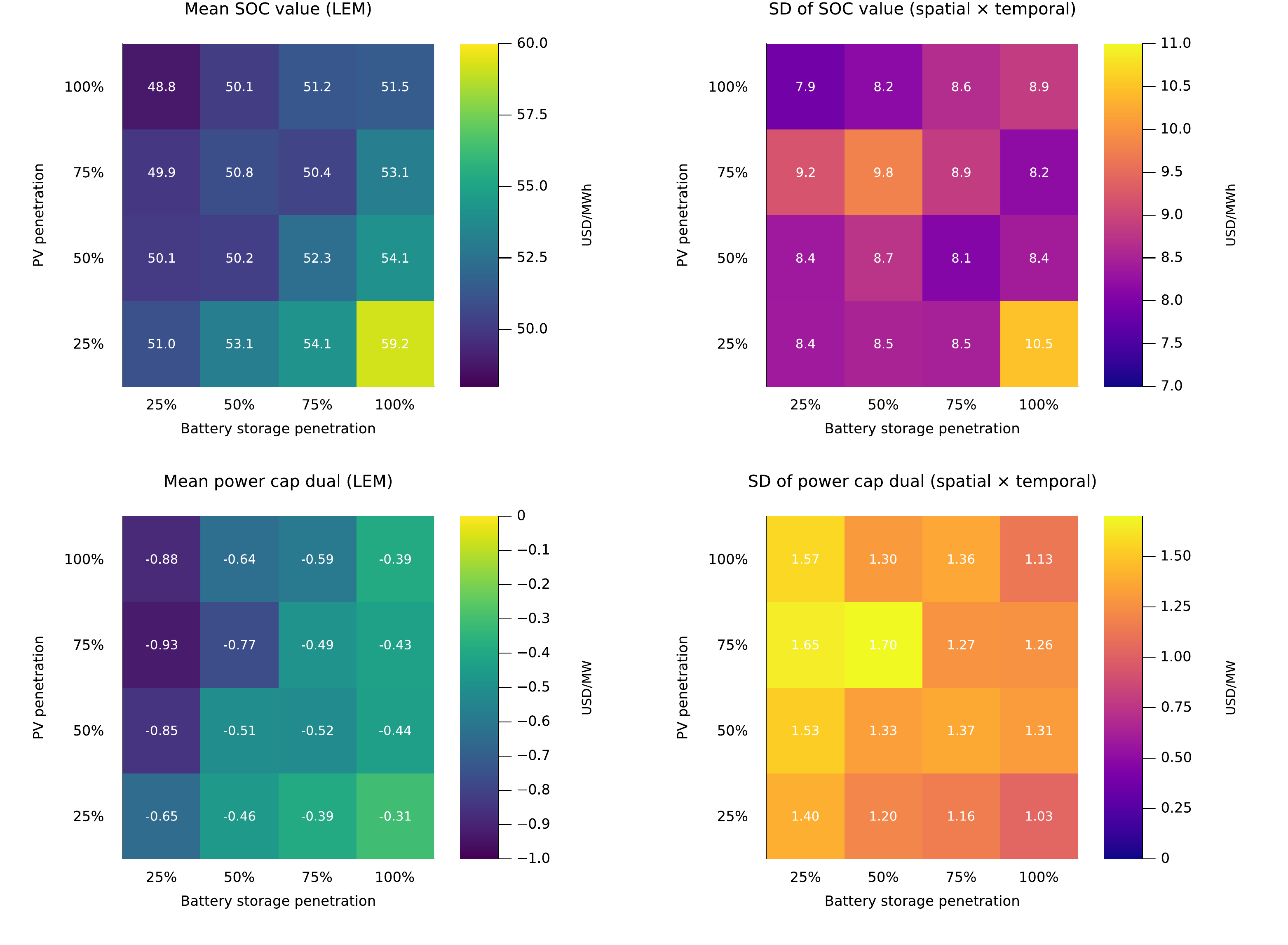}
  \caption{Sensitivity analysis of the SOC dual-based storage value.}
  \label{fig:soc_dual_sensitivity}
\end{figure}

Both the SOC value and power cap duals are reported as daily means and averaged over the entire network as shown in figure \ref{fig:soc_dual_sensitivity}. We find that the SOC duals decrease with PV penetration but increase with BS penetration -- essentially serving as a measure of the imbalance between local generation available and storage. At low PV and high BS, many batteries competing for scarce local generation makes each MWh of stored battery energy highly valuable. At high PV and low BS, abundant generation but little storage means the SOC constraint is less binding. We observe that the average SOC duals across the penetration space are significantly higher than the average wholesale real-time market LMP for the Boston area in August 2025, which was \$41.61/MWh \cite{isone2025lmpindices}. This premium reflects the scarcity value of batteries embedded in d-LMPs that wholesale prices do not capture, arising from factors like losses and voltage constraints as well as services like network support, congestion relief, and localized flexibility. Thus, distribution-level storage can consistently capture more value than pure wholesale arbitrage, provided that local generation is sufficient to maintain utilization -- making the LEM more attractive for battery owners than other schemes like RTP. The standard deviation across nodes and timesteps is quite stable (8-10 \$/MWh) throughout the penetration space, representing roughly 16-18\% variation around the mean. The highest volatility occurs at 25\% PV + 100\% BS, implying that mismatched DER configurations (many batteries, few generators) create more variable market conditions. Such stable levels of SOC value volatility also confirm there are genuine spatial and temporal arbitrage opportunities for batteries, especially when there is more imbalance between solar and battery penetrations. Variations in the SOC value can also provide valuable information for different stakeholders. For instance, rising SOC duals would indicate that the scarce resource is generation rather than storage, so further BS installations would yield diminishing returns since it would further shorten the relative shortage of PV and make it challenging to charge the batteries to the desired level. This could inform policymakers to customize DER subsidies or incentive programs based on local generation-to-storage ratios, and guide grid operators before further storage interconnections. Battery owners and developers could similarly use this information to prioritize co-locating storage projects with PV. Finally, this stored energy value can also be interpreted as the LEM's direct battery dispatch signal, with high values incentivizing aggressive charging and low values favoring idle operation or discharging. This highlights how the LEM is able to effectively coordinate decentralized batteries using distribution-level price signals, making it superior to wholesale RTP-based alternatives.

The power cap duals reflect the shadow prices on the battery power rating constraints, i.e., the maximum charging or discharging rate at each node. All the mean values are negative implying that relaxing these constraints would lower system costs. The relatively small magnitudes indicate that the battery power capacity limits are only active or binding for a small portion of times and nodes, likely during large load spikes or ramps (up or down) in PV output (sunrise, sunset, cloud cover, etc.). Outside those episodes, batteries operate well within their power limits and these duals are zero. Thus, battery energy capacity seems to be the more important limiting factor rather than rated power capacity in general, apart from some occasional high-value ramping episodes. We see a clear trend that the power cap duals generally become more binding (more negative) as we increase PV penetration and less binding with increasing BS penetration. This makes intuitive sense -- abundant solar generation creates large, fast ramps that the battery fleet needs to absorb, while larger BS penetrations mean that more batteries share the ramp burden, so fewer individual units are rate-constrained. The high standard deviations relative to the mean also indicate significant spatial and temporal volatility mainly driven by rapid ramping in load and PV. As expected intuitively, this volatility also rises with PV penetration and falls with BS penetration.
 
\subsection{Discussion}
The results show a clear ranking across the three market designs: the LEM performs best, followed by FT, while RTP performs worst. The LEM yields the lowest system cost and the lowest customer cost. However, the operator cost under the LEM is higher than under FT. This may discourage LEM adoption at first, but the operator revenue loss (41\% relative to FT) is outweighed by the customer cost reductions (33\% relative to FT), resulting in a net system cost saving of 22\% over FT. Even though operators still earn revenue under the LEM, other approaches like fixed rebates could be used to share these efficiency gains more evenly between the customer and the operator.

The standard deviations reported in \cref{tab:monthly_costs_breakdown} also show that the LEM operator and customer costs are more sensitive to uncertainty than the benchmark designs. The LEM has a standard deviation of 7{,}623~\$/month in operator cost and 7{,}747~\$/month in customer cost, compared with much smaller variations under FT and RTP. This is consistent with the role of forecast-based coordination in the LEM: because battery dispatch responds more actively to expected feeder conditions, realized outcomes depend more strongly on the realized PV path. By contrast, the benchmark designs provide weaker coordination signals and therefore produce more stable, but less efficient, outcomes. In other words, the LEM delivers better average performance, but it is also more exposed to uncertainty. Considering lost load, the LEM has essentially zero variation in lost-load cost across all OU scenarios, while RTP is strongly exposed to reliability deterioration under unfavorable PV realizations. As lost load cost is a part of system cost, RTP's high standard deviation translates directly to system cost.

It is important to note that the LMP series used in all simulations does not contain any negative prices. In real renewables-rich power systems, however, negative prices do occur. If such price periods were included, the LEM would probably do even better since it explicitly coordinates all DERs and is therefore better positioned to respond efficiently to negative-price signals. \cref{fig:energy_balance_50_50} shows that all markets serve similar demand. Also, the DER energy flows are rather small in comparison to overall demand. However, the contrast in costs shows how relatively small energy flows from DERs at the right place and time can have a big impact on system efficiency. Poorly coordinated DERs, as in RTP, are not able to avoid lost load and therefore suffer from high costs, even though they have a similar energy flow level as the LEM.

In \cref{System_costs_bs_variable}, once batteries are available, the LEM reduces lost load much more effectively than the benchmark designs, and from a battery penetration of 10\% onward it eliminates lost load entirely. As there are high lost load cost for all markets without any battery penetration, it indicates that the feeder is heavily stressed in the absence of flexibility. RTP lost load cost decreases initially to a minimum at 10\% battery penetration, but then starts to increase with increasing battery penetration. This is due to the egoistic battery operation that does not consider grid state but only nodal profit maximization. Thus, a high penetration of egoistical battery operation can badly influence a distribution grid under RTP. FT can not fully eliminate lost load, which even increases slightly as battery penetration grows, because it remains blind to the location of congestion and therefore cannot coordinate flexibility where it is most valuable. The comparison between lost load cost and wholesale energy cost also suggests that the LEM first uses scarce battery flexibility to protect reliability. In \cref{System_costs_bs_variable}, lost load cost falls sharply at low battery penetration, while wholesale energy import cost slightly increases for low battery penetration. This pattern is consistent with the LEM objective: because the penalty on lost load is much larger than the cost of wholesale energy imports, the optimization prioritizes avoiding lost load before using remaining flexibility for wholesale energy cost reduction. Once lost load has been eliminated, additional battery capacity can increasingly be used to reduce wholesale energy cost.

The heatmaps in \cref{fig:heatmap_customer_cost} and \cref{fig:heatmap_operator_cost} further suggest that the LEM uses available DER capacity more effectively than the benchmark designs, as long as the system's PV penetration is above a low threshold (15--20\%). In the LEM, relatively small amounts of DER capacity already produce large changes in operator and customer costs, whereas the benchmark designs either improve more gradually as DER penetration rises, or can even degrade. For instance, operator costs under RTP increase with battery penetration. This indicates that under d-LMP-based coordination, DERs are used more effectively from the system perspective, while under FT a larger installed capacity is needed to achieve similar improvements.

The battery valuation results provide several insights. First, the value of added storage in an LEM is much larger when the feeder starts from low battery penetration, in terms of both private and system-wide benefits. Second, location matters. The best node from a private perspective may not always be the best node from a system perspective. This creates a potential system-private gap in storage siting incentives. In practice, this could justify targeted subsidies or other support mechanisms for batteries installed at locations where the system benefit is high but the private incentive is too weak (or even negative). At higher battery penetration of about 10\% to 20\%, the system value of additional storage becomes very small, as seen in \cref{fig:systemMV_node61}. This is consistent with the literature about other distribution market designs, as well as with the earlier reliability results. Once the lost load is already eliminated, the remaining system benefit of additional storage comes mainly from reduced wholesale energy imports and congestion management, which are smaller value streams than avoided curtailment in this setup. At that point, storage placement becomes more sensitive to local network conditions and nodal price patterns.


In contrast to the sharp decrease in the system marginal value of batteries at high battery penetration, the private value can remain meaningful at selected locations. \cref{tab:marginal_value_20PV} shows the average marginal values and standard deviations for all four tested cases. Interestingly, the standard deviation of the private marginal value for 10\% and 20\% battery penetration is quite high with 5.8 \$/kWh/day and 4.4 \$/kWh/day, respectively. This means that battery placement is especially crucial if there is an existing but low battery penetration. It is worth noting that the mean private marginal values for all tested penetration levels, ranging from 0.3 \$/kWh/day for the 40\% battery penetration to 6.7 \$/kWh/day for 10\% battery penetration, are highly lucrative, as a lithium-ion battery cell costs around 60 \$/kWh \cite{owid_battery_cell_prices_page}. The volatility of private marginal value based on different penetrations is visible in \cref{fig:privateMV_node61} for node 61. Only for private marginal value at low battery penetration levels, below 10\%, is there some consistency. Afterwards, no clear trend is visible, with some positive outliers, for example, at 0\% PV / 50\% BS or at 35\% PV / 30\% BS. But there are also outliers in the direction of decreasing marginal value when battery capacity is increased at node 61 under specific DER penetrations, for example at 40\% PV / 60\% BS or 5\% PV / 55\% BS. This behavior can be explained by different DER coordination needs and therefore different bottlenecks at different penetration levels.

\subsection{Future work}
There are several potential areas to extend and build upon this study. Our analysis only focuses on operational performance and variable costs rather than long-run investment decisions, which would require considering the fixed capital costs for infrastructure. Such long-term studies would also entail modified OU-scenario generation to capture uncertainty and trends over much longer time horizons. We also made several assumptions that can be relaxed. All storage systems have the same \(C\)-rate, and all agents are assumed to be rational and price-taking. In addition, wholesale prices are represented exogenously at the PCC rather than through full transmission-distribution co-optimization, and reliability is represented only through VOLL-priced lost load rather than through a broader resilience framework. Also, only one random distribution of 20 PV plants and 20 batteries is evaluated. The OU process is also limited to one parameter setting. These modeling choices and simplifications keep the problem tractable and make it possible to isolate the effects of market design, uncertainty, and storage placement, but they also limit the direct interpretation of the results for real-world implementation. Refining these assumptions and increasing model complexity would bring the study closer to practical application.

\section{Conclusion}

This paper evaluated a d-LMP-settled local electricity market (LEM) on the IEEE-123 feeder under PV forecast uncertainty, modeled through an Ornstein-Uhlenbeck process and implemented in a receding-horizon framework. The analysis compared the LEM with two conventional retail market designs, flat tariff + net energy metering and real-time pricing with net energy billing, under identical DER portfolios, full AC-feasible network evaluation, and the same realized uncertainty trajectories. The results show that, even under explicitly modeled PV forecast uncertainty, the LEM outperforms the benchmark designs in terms of reliability and average system cost. In the stressed feeder considered here, the LEM eliminates lost load, while FT reduces but does not eliminate lost load and RTP performs worst. The LEM also yields the lowest system and customer cost once sufficient battery capacity is available, indicating that nodal, network-aware coordination uses storage more effectively than tariff-based alternatives that provide only uniform or time-varying price signals. While the LEM performs best on average, its operator and customer costs vary more strongly across PV forecast realizations than those under RTP and FT. This reflects the fact that forecast-based coordination makes realized market outcomes more sensitive to the realized PV path. Through extensive simulations over a range of PV and BS penetrations, we find that the LEM vastly outperforms the other markets above a minimum PV penetration threshold (around 15--20\%), but may result in higher costs for customers at very low PV levels. This indicates that LEM implementations should also be combined with driving PV adoption, and that PV adoption must be prioritized before increasing BS deployment.

We also developed methods to accurately assess the value of both new and existing storage capacity under the LEM. We show that the value of additional battery capacity in a LEM is strongly location-dependent and declines with increasing baseline battery and PV penetration. When initial battery penetration is low, additional storage creates substantial private and system value. As battery penetration rises, marginal system value declines, while private value can remain meaningful at selected nodes. Moreover, the best location from a private perspective does not always coincide with the best location from a system perspective, indicating a role for targeted policy support or compensation mechanisms. On the other hand, the scarcity value of stored energy in the existing battery fleet value rises when battery penetration outpaces PV and falls when solar PV is abundant relative to batteries. 
Overall, the results suggest that LEMs can improve the operation of DER-rich distribution feeders under forecast uncertainty by coordinating storage where and when it is most valuable.  We also present an alternative approach to value storage based on our optimization model, mirroring dual-based electricity pricing. Accurately understanding the value of installed and additional storage is crucial for policymakers, grid operators, and battery owners and developers, to more optimally plan, prioritize, and invest. Future work could extend the framework in several directions. One natural extension would be to examine targeted battery-support schemes that encourage siting at system-beneficial locations by leveraging our optimization framework. Simulations over longer time horizons could capture long-term trends as well as other seasons, beyond the summer months considered here. Similar to the OU modeling approach used for solar radiation, uncertainty in other key inputs (e.g., prices, load profiles) can be considered using appropriate stochastic processes. Finally, it would be interesting to include other emerging types of DERs like heat pumps and electric vehicles in these case studies.

\section{Acknowledgments}

This research was supported by the MIT Energy Initiative.

\section{CRediT authorship contribution statement}

\textbf{Peer Brigger}: Conceptualization, Data curation, Formal analysis, Investigation, Methodology, Resources, Software, Validation, Visualization, Writing – original draft, Writing – review and editing. \textbf{Vineet Jagadeesan Nair}: Conceptualization, Methodology, Resources, Software, Visualization, Supervision, Writing – original draft, Writing – review and editing. \textbf{Anuradha M. Annaswamy}: Conceptualization, Project administration, Funding acquisition, Supervision, Writing – review and editing.

\section{Declaration of competing interests}
The authors declare that they have no known competing financial interests or personal relationships that could have appeared to influence the work reported in this paper.

\section{Data availability}
Detailed descriptions of the model equations, variables, parameters, major input and output data, and corresponding data sources are available in the main text. Any additional data needed will be made available on request.

\bibliographystyle{elsarticle-num}

\bibliography{cas-refs}

@techreport{IEA_Electricity_2025,
  title       = {Electricity 2025},
  institution = {{International Energy Agency}},
  year        = {2025},
  note        = {Executive summary and Demand chapter}
}

@article{Reuters2024ISOELeGrowth,
  title   = {New England grid prepared for summer electricity demand, operator says},
  author  = {Reuters},
  year    = {2024},
  url     = {https://www.reuters.com/world/us/},
  note    = {Reported Nov.\ 2024; Accessed 2025-09-23}
}

@misc{ISO-NE2024CELTOverview,
  title   = {2024 CELT: Forecast Report of Capacity, Energy, Loads, and Transmission --- Overview},
  author  = {{ISO New England}},
  year    = {2024},
  url     = {https://www.iso-ne.com/static-assets/documents/2024/05/2024-celt-forecast-overview.pdf},
  note    = {Accessed 2025-09-23}
}

@misc{isone2025lmpindices,
  author       = {{ISO New England}},
  title        = {Monthly {LMP} Indices},
  year         = {2025},
  howpublished = {ISO New England ISOExpress},
  url          = {https://www.iso-ne.com/isoexpress/web/reports/pricing/-/tree/monthly-lmp-indices},
  note         = {Simple monthly averages of hourly Hub and zonal locational marginal prices; August 2025 data accessed from 2025 annual CSV download}
}

@misc{ISO-NE2024Press,
  title   = {2024 CELT Forecast Press/Newswire Summary},
  author  = {{ISO New England}},
  year    = {2024},
  howpublished = {\url{https://isonewswire.com/2024/05/23/celt-2024-expanded-economic-model-informs-long-term-forecast/}},
  note    = {Baseline electrification drivers; Accessed 2025-09-23}
}

@misc{Powell2026LinkedIn,
  author       = {Mary Powell},
  title        = {Local Solar and Storage Will Offset Grid Pressure from AI and Data Centers},
  year         = {2026},
  month        = mar,
  howpublished = {LinkedIn article},
  note         = {Published March 4, 2026},
  url          = {https://www.linkedin.com/pulse/local-solar-storage-offset-grid-pressure-from-ai-data-mary-powell-8xnae/}
}

@techreport{IEA-PVPS-USA-2023,
  title   = {National Survey Report of PV Power Applications in the United States 2023},
  author  = {{IEA PVPS Task 1}},
  year    = {2024},
  url     = {https://iea-pvps.org/national_survey/national-survey-report-of-pv-power-applications-in-the-usa-2023/},
  note    = {Table 1 shows small-scale (decentralized) PV: 47{,}704 MW\,(AC) cumulative; 7{,}876 MW\,(AC) added in 2023.}
}

@article{Reuters2025CapacityMix,
  title   = {Charting the projected U.S. power capacity mix through 2035},
  author  = {Reuters Graphics},
  year    = {2025},
  url     = {https://www.reuters.com/business/energy/charting-projected-us-power-capacity-mix-through-2035-2025-08-21/},
  note    = {Summarizes EIA projections; Accessed 2025-09-23}
}

@techreport{NERC2024SRA,
  title  = {2024 Summer Reliability Assessment},
  author = {{North American Electric Reliability Corporation (NERC)}},
  year   = {2024},
  url    = {https://www.nerc.com/globalassets/programs/rapa/ra/nerc_sra_2024.pdf}
}

@article{Lamont2013StorageValue,
  title   = {Assessing the Economic Value and Optimal Structure of Large-Scale Electricity Storage},
  author  = {Lamont, Alan D.},
  journal = {IEEE Transactions on Power Systems},
  volume  = {28},
  number  = {2},
  pages   = {911--921},
  year    = {2013},
  doi     = {10.1109/TPWRS.2012.2218135}
}

@article{Sisternes2016StorageDecarb,
  title   = {The Value of Energy Storage in Decarbonizing the Electricity Sector},
  author  = {de Sisternes, Fernando J. and Jenkins, Jesse D. and Botterud, Audun},
  journal = {Applied Energy},
  year    = {2016},
  volume  = {175},
  pages   = {368--379}
}

@inproceedings{Engelman2025LMVOutage,
  title     = {Locational Marginal Value of Storage Under Risk of Supply-Side Disruption},
  author    = {Engelman Lado, Nathan and Khorramfar, Rahman and Amin, Saurabh},
  booktitle = {Proceedings of the American Control Conference (ACC)},
  year      = {2025},
  address   = {Denver, CO, USA},
  month     = jul
}

@inproceedings{Bose2015LMVStorage,
  title     = {Variability and the Locational Marginal Value of Energy Storage},
  author    = {Bose, Subhonmesh and Bitar, Eilyan},
  booktitle = {Proceedings of the 53rd {IEEE} Conference on Decision and Control ({CDC})},
  pages     = {3259--3265},
  year      = {2014},
  address   = {Los Angeles, CA, USA},
  doi       = {10.1109/CDC.2014.7039893}
}

@misc{UtilityDive2025Capex,
  title   = {Electric utilities will invest more than \$1.1{T} by 2030 to meet demand growth: {EEI}},
  author  = {Walton, Robert},
  year    = {2025},
  month   = jul,
  url     = {https://www.utilitydive.com/news/electric-utilities-will-invest-more-than-11t-by-2030-to-meet-demand-growt/753783/},
  note    = {Utility Dive, July 23, 2025}
}

@techreport{EEI2024CapexPage,
  title       = {2024 {Financial Review}: Annual Report of the {U.S.} Investor-Owned Electric Utility Industry},
  author      = {{Edison Electric Institute}},
  institution = {{Edison Electric Institute}},
  year        = {2025},
  url         = {https://www.eei.org/-/media/Project/EEI/Documents/Issues-and-Policy/Finance-And-Tax/Financial_Review/FinancialReview_2024.pdf}
}

@techreport{Hinchberger2024CEEPR,
  title       = {The Efficiency of Dynamic Energy Prices},
  author      = {Hinchberger, Andrew and Jacobsen, Mark R. and Knittel, Christopher R.
                 and Sallee, James M. and {van Benthem}, Arthur A.},
  institution = {{MIT Center for Energy and Environmental Policy Research (CEEPR)}},
  year        = {2024},
  number      = {CEEPR WP-2024-15},
  url         = {https://ceepr.mit.edu/wp-content/uploads/2024/10/MIT-CEEPR-WP-2024-15.pdf},
  note        = {Working paper}
}

@techreport{CEEPR2022DynamicTariffs,
  title   = {Electricity Retail Rate Design in a 
Decarbonizing Economy: An Analysis of 
Time-of-Use and Critical Peak Pricing},
  author  = {{MIT CEEPR}},
  year    = {2022},
  url     = {https://ceepr.mit.edu/}
}

@article{Tan2022LMPReview,
  title   = {Extensions of the Locational Marginal Price Theory in
             Evolving Power Systems: A Review},
  author  = {Tan, Zhenfei and Cheng, Tong and Liu, Yuchen and Zhong, Haiwang},
  journal = {IET Generation, Transmission \& Distribution},
  volume  = {16},
  number  = {7},
  pages   = {1277--1291},
  year    = {2022},
  doi     = {10.1049/gtd2.12381}
}

@article{Wang2023d-LMPVision,
  title   = {Distribution Locational Marginal Pricing of Competitive Markets in Active Distribution Networks: Models, Solutions, Applications, and Visions},
  author  = {Wang, Y. and Li, Z. and others},
  journal = {Renewable and Sustainable Energy Reviews},
  volume  = {171},
  pages   = {113017},
  year    = {2023},
  doi     = {10.1016/j.rser.2022.113017}
}

@article{Domenech2024d-LMPDecomp,
  title   = {Towards Distributed Energy Markets: Accurate and Intuitive
             {DLMP} Decomposition},
  author  = {{Bas Domenech}, Carmen and Naughton, James and Riaz, Shariq
             and Mancarella, Pierluigi},
  journal = {IEEE Transactions on Energy Markets, Policy and Regulation},
  year    = {2024},
  doi     = {10.1109/TEMPR.2024.3360996}
}

@article{NairAnnaswamy2023,
  title   = {A Hierarchical Local Electricity Market for a {DER}-rich Grid Edge},
  author  = {Nair, Vineet Jagadeesan and Venkataramanan, Venkatesh
             and Haider, Rabab and Annaswamy, Anuradha M.},
  journal = {IEEE Transactions on Smart Grid},
  volume  = {14},
  number  = {4},
  pages   = {3220--3232},
  year    = {2023},
  doi     = {10.1109/TSG.2022.3174036}
}

@article{haider2021,
  title   = {Reinventing the Utility for Distributed Energy Resources: A Proposal for Local Retail Electricity Markets},
  author  = {Haider, Rabab and D'Achiardi, David and Venkataramanan, Venkatesh and Annaswamy, Anuradha M.},
  journal = {Advances in Applied Energy},
  volume  = {2},
  pages   = {100026},
  year    = {2021},
  doi     = {10.1016/j.adapen.2021.100026}
}

@article{Li2014EVCS_d-LMP,
  title   = {Distribution Locational Marginal Pricing for Optimal Electric Vehicle Charging Management},
  author  = {R. Li, Q. Wu and S. S. Oren},
  journal = {IEEE Transactions on Power Systems},
  volume  = {29},
  number  = {1},
  pages   = {203--211},
  year    = {2014},
  doi     = {10.1109/TPWRS.2013.2278952}
}

@article{Potter2023ReactivePowerMarket,
  title   = {A reactive power market for the future grid},
  author  = {Potter, Adam and Haider, Rabab and Ferro, Giulio and Robba, Michela and Annaswamy, Anuradha M.},
  journal = {Advances in Applied Energy},
  volume  = {9},
  pages   = {100114},
  year    = {2023},
  month   = feb,
  doi     = {10.1016/j.adapen.2022.100114},
  url     = {https://doi.org/10.1016/j.adapen.2022.100114}
}

@article{Krishna2024LMPBESS,
  title   = {Locational marginal price based scheduling strategy for
             effective utilization of battery energy storage in
             {PV} integrated distribution system},
  author  = {Krishna, Swathi and Deepak, M. and Sunitha, R.},
  journal = {Journal of Energy Storage},
  volume  = {94},
  pages   = {112102},
  year    = {2024},
  doi     = {10.1016/j.est.2024.112102}
}

@article{FaruquiSergici2010,
  title   = {Household Response to Dynamic Pricing of Electricity: A Survey of 15 Experiments},
  author  = {Faruqui, Ahmad and Sergici, Sanem},
  journal = {Journal of Regulatory Economics},
  volume  = {38},
  number  = {2},
  pages   = {193--225},
  year    = {2010},
  doi     = {10.1007/s11149-010-9127-4}
}

@misc{CPUC_NEM_NBT,
  title   = {Net Energy Metering and Net Billing},
  author  = {{California Public Utilities Commission}},
  year    = {2025},
  note    = {Explains that NEM exports receive bill credits at retail rates; summarizes transition to Net Billing},
  url     = {https://www.cpuc.ca.gov/industries-and-topics/electrical-energy/demand-side-management/customer-generation/net-energy-metering-and-net-billing}
}

@techreport{NREL_DG_Comp,
  title   = {Grid-connected Distributed Generation: Compensation Mechanisms},
  author  = {O. Zinaman and others},
  institution = {{National Renewable Energy Laboratory (NREL)}},
  number  = {NREL/TP-6A20-68469},
  year    = {2017},
  url     = {https://www.nrel.gov/docs/fy18osti/68469.pdf}
}

@article{Zhao2018d-LMPUncertainty,
  title   = {Distribution Locational Marginal Pricing Under Uncertainty
             Considering Coordination of Distribution and Wholesale Markets},
  author  = {Zhao, Zongzheng and Liu, Yixin and Guo, Li
             and Bai, Linquan and Wang, Chengshan},
  journal = {IEEE Transactions on Smart Grid},
  volume  = {14},
  number  = {2},
  pages   = {1590--1606},
  year    = {2023},
  doi     = {10.1109/TSG.2022.3200704}
}

@article{Robustd-LMP2022,
  title   = {Bi-Level Robust Optimization for Distribution System With
             Multiple Microgrids Considering Uncertainty Distribution
             Locational Marginal Price},
  author  = {Wang, Lingling and Zhu, Zean and Jiang, Chuanwen and Li, Zuyi},
  journal = {IEEE Transactions on Smart Grid},
  volume  = {12},
  number  = {2},
  pages   = {1104--1117},
  year    = {2021},
  doi     = {10.1109/TSG.2020.3037556}
}

@article{MPCMILP2022,
  title   = {Model predictive control for optimal power flow in
             grid-connected unbalanced microgrids},
  author  = {Erazo-Caicedo, David and Mojica-Nava, Eduardo
             and Revelo-Fuelag{\'a}n, Javier},
  journal = {Electric Power Systems Research},
  volume  = {209},
  pages   = {108000},
  year    = {2022},
  doi     = {10.1016/j.epsr.2022.108000}
}

@misc{SocialWelfareMPC2020,
  title        = {A Distributed Economic Model Predictive Control Design for a Transactive Energy Market Platform in Lebanon, NH},
  author       = {Muhanji, Steffi Olesi and Golding, Samuel and Montgomery, Tad and Below, Clifton and Farid, Amro M.},
  year         = {2020},
  eprint       = {2012.04058},
  archivePrefix= {arXiv},
  primaryClass = {eess.SY},
  url          = {https://arxiv.org/abs/2012.04058}
}

@article{CarteaFigueroa2005OU,
  title   = {Pricing in Electricity Markets: A Mean-Reverting Jump-Diffusion Model with Seasonality},
  author  = {Cartea, {\'A}lvaro and Figueroa, Marcelo G.},
  journal = {Applied Mathematical Finance},
  year    = {2005},
  volume  = {12},
  number  = {4},
  pages   = {313--335},
  doi     = {10.1080/13504860500151032}
}

@techreport{MISO_Scarcity_VOLL2024,
  title   = {Scarcity Pricing White Paper: Value of Lost Load and ORDC},
  author  = {{Midcontinent Independent System Operator (MISO)}},
  year    = {2024},
  note    = {Discusses raising VOLL to \$10,000/MWh as a price cap and administrative price},
  url     = {https://cdn.misoenergy.org/20240418%20MSC%20Item%2004d%20Scarcity%20Pricing%20White%20Paper%20VOLL%20and%20ORDC632355.pdf}
}

@misc{NexansACSRData,
  title        = {{ACSR} Electrical Data},
  author       = {{Nexans}},
  year         = {n.d.},
  howpublished = {\url{https://www.nexans.us/.rest/eservice/dam/v1/file/224198/ACSR_Electrical_Data.pdf}},
  note         = {Datasheet with typical ampacity values (e.g., Raven, Penguin, Linnet, Hawk, Pelican)
                  under standard assumptions (75\(^\circ\)C conductor temperature, 25\(^\circ\)C ambient,
                  $\sim$2\,ft/s crosswind, sun)}
}

@book{BoydVandenberghe2004,
  title     = {Convex Optimization},
  author    = {Boyd, Stephen and Vandenberghe, Lieven},
  year      = {2004},
  publisher = {Cambridge University Press},
  url       = {https://web.stanford.edu/~boyd/cvxbook/}
}

@misc{Eversource_RatesTariffs,
  title   = {Rates and Tariffs -- Eversource Energy},
  author  = {{Eversource Energy}},
  year    = {2025},
  howpublished = {Residential Rate and Tariff Information},
  note    = {Accessed October 2025. Provides current retail electricity rate structures and billing options.},
  url     = {https://www.eversource.com/residential/account-billing/manage-bill/about-your-bill/rates-tariffs}
}

@inproceedings{NairLocalRetail2023,
  title     = {Local Retail Electricity Markets for Distribution Grid Services},
  author    = {Nair, Vineet Jagadeesan and Annaswamy, Anuradha M.},
  booktitle = {2023 IEEE Conference on Control Technology and Applications (CCTA)},
  pages     = {32--39},
  year      = {2023},
  publisher = {IEEE},
  doi       = {10.1109/CCTA54093.2023.10253051}
}

@article{Bai2018d-LMPVoltage,
  title   = {Distribution Locational Marginal Pricing ({DLMP}) for Congestion Management and Voltage Support},
  author  = {L. Bai and J. Wang and C. Wang and C. Chen and F. Li},
  journal = {{IEEE} Transactions on Power Systems},
  volume  = {33},
  number  = {4},
  pages   = {4061--4073},
  year    = {2018},
  doi     = {10.1109/TPWRS.2017.2767632}
}

@misc{owid_battery_cell_prices_page,
  author       = {Hannah Ritchie and Pablo Rosado and Max Roser},
  title        = {Data Page: Lithium-ion battery cell prices by chemistry},
  year         = {2023},
  organization = {Our World in Data},
  note         = {Data adapted from Benchmark Mineral Intelligence. Archived on March 4, 2026. Retrieved March 19, 2026},
  url          = {https://archive.ourworldindata.org/20260304-094028/grapher/average-battery-cell-price.html}
}

@misc{ISO-NE_FiveMinuteLMPs,
  title        = {Final Real-Time Five-Minute LMPs},
  author       = {{ISO New England}},
  year         = {2026},
  note         = {ISO Express pricing report providing final approved locational marginal prices for each five-minute interval of the operating day},
  url          = {https://www.iso-ne.com/isoexpress/web/reports/pricing/-/tree/lmps-rt-five-minute-final},
  urldate      = {2026-05-14}
}

\end{document}